\documentclass{jpp}
\usepackage{graphicx,xcolor}
\usepackage{epstopdf, epsfig}
\usepackage{hyperref}
\newcommand{\red}[1]{}
\def\be{\begin{equation}}
\def\ee{\end{equation}}
\def\ba{\begin{eqnarray}}
\def\ea{\end{eqnarray}}
\def\uu{{\bf u}}
\def\bb{{\bf B}}
\def\bbs{{\bf b}}
\def\kk{{\bf k}}
\def\pp{{\bf p}}
\def\qq{{\bf q}}
\def\kpa{k_{\parallel}}
\def\ppa{p_{\parallel}}
\def\qpa{q_{\parallel}}
\def\kpe{k_{\perp}}
\def\ppe{p_{\perp}}
\def\qpe{q_{\perp}}
\def\ppeb{{\bf p_{\perp}}}
\def\qpeb{{\bf q_{\perp}}}
\def\dd{\mathrm{d}}
\def\aks{c^s_k}
\def\akss{c^{s'}_{k'}}
\def\aksss{c^{s''}_{k''}}
\def\akssss{c^{s'''}_{k'''}}
\def\RR{{\mathbb{R}}}
\def\aksp{c^{s_{p}}_{p}}
\def\aksq{c^{s_{q}}_{q}}
\def\Okpq{\Omega_{k,pq}}
\def\Okpqp{\Omega_{k',pq}}
\def\Okpqpp{\Omega_{k'',pq}}

\def\del{\delta_{k,pq}}
\def\delp{\delta_{k',pq}}
\def\delpp{\delta_{k'',pq}}

\def\Aks{A^s_k}
\def\Akss{A^{s'}_{k'}}

\shorttitle{Fast magneto-acoustic wave turbulence}
\shortauthor{S\'ebastien Galtier}

\title{Fast magneto-acoustic wave turbulence and the Iroshnikov-Kraichnan spectrum}
\author{S\'ebastien Galtier \aff{1,2,3}
\corresp{\email{sebastien.galtier@lpp.polytechnique.fr}}}

\affiliation{\aff{1}Laboratoire de Physique des Plasmas, \'Ecole polytechnique, F-91128 Palaiseau Cedex, France
\aff{2}Universit\'e Paris-Saclay, IPP, CNRS, Observatoire Paris-Meudon, France
\aff{3}Institut universitaire de France}

\begin{document}

\maketitle

\begin{abstract}
An analytical theory of wave turbulence is developed for pure compressible magnetohydrodynamics in the small $\beta$ limit. In contrast to previous works where the multiple scale method was not mentioned and slow magneto-acoustic waves were included, I present here a theory for fast magneto-acoustic waves only for which an asymptotic closure is possible in three dimensions. I introduce the compressible Els\"asser fields (canonical variables) and show their linear relationship with the mass density and the compressible velocity. The kinetic equations of wave turbulence for three-wave interactions are obtained and the detailed conservation is shown for the two invariants, energy and momentum (cross-helicity). An exact stationary solution (Kolmogorov-Zakharov spectrum) exists only for the energy. I find a $k^{-3/2}$ energy spectrum compatible with the Iroshnikov-Kraichnan (IK) phenomenological prediction; this leads to a mass density spectrum with the same scaling. Despite the presence of a relatively strong uniform magnetic field, this turbulence is characterized by an energy spectrum with a power index that is independent of the angular direction; its amplitude, however, shows an angular dependence. I prove the existence of the IK solution using the locality condition, show that the energy flux is positive and hence the cascade direct, and find the Kolmogorov constant. This theory offers a plausible explanation for recent observations in the solar wind at small $\beta$ where isotropic spectra with a $-3/2$ power law index are found and associated with fast magneto-acoustic waves. This theory may also be used to explain the IK spectrum often observed near the Sun. Besides, it provides a rigorous theoretical basis for the well-known phenomenological IK spectrum, \red{which coincides with the Zakharov-Sagdeev spectrum for acoustic wave turbulence}.
\end{abstract}

\section{Introduction}\label{Sec1}
\subsection{Solar wind turbulence}
The solar wind has been studied for many years and despite its proximity and the fact that spacecraft have been launched to discover its turbulent properties, several questions still remain open \citep{Goldstein1999,Bruno2013,Sahraoui2020}. However, it would be wrong to think that solar wind turbulence is not understood at all as we have made significant progress in this area over the last few decades. For example, the existence of a finite inertial range for the applicability of magnetohydrodynamics (MHD) is now well established, as is the fact that this turbulence consists of Alfv\'en waves and is anisotropic \citep{Matthaeus2021}. To interpret the anisotropy at 1\,AU, the critical balance phenomenology \citep{Higdon1984,Goldreich1995,Oughton2020} is often used. This simple model of strong incompressible MHD turbulence predicts, for the energy spectrum in the direction transverse to the local mean magnetic field, a power law index $\alpha = -5/3$, which is often observed for the magnetic fluctuation spectrum \citep{Podesta2007}. However, this interpretation has limitations since the power law index observed for the velocity fluctuation spectrum is $\alpha = -3/2$. If we want to better understand sub-MHD scales, it is also recognized that the MHD approximation must be improved with, in particular, the introduction of new nonlinear effects such as the Hall effect \citep{Galtier2006,Passot2019}. 
The higher resolution observations provided by Cluster/ESA \citep{Bale2005,Kiyani2015} have led us to propose new theories for plasma turbulence. As far as we are concerned, we can mention the generalization of the exact (MHD) Kolmogorov law \citep{K41,Politano1998b,Galtier2008b} for compressible turbulence, first in the case of isothermal hydrodynamics \citep{Galtier2011} and then to MHD (with different closures and/or scales description) \citep{Banerjee2013,Andres2018,Ferrand2021,Simon2022}. Second, the extensive use of these (compressible) laws as a solar wind model has led to a better estimate of the turbulent transfer and thus of the local heating, although we still do not know precisely by what mechanism this small scale heating occurs \citep{Sorriso2007,Osman2011,Banerjee2016,Hadid2017,Bandyopadhyay2020,Marino2023}. 

The most remarkable recent solar wind observations come from PSP/NASA (Parker Solar Probe): they reveal a universal behaviour of the solar wind near the Sun ($\sim 0.1$\,AU) with identical power law indices $\alpha = -3/2$ for the velocity and magnetic fluctuation spectra \citep{Chen2020,Shi2021,Zhao2022}. The mass density spectrum measured by PSP is also roughly compatible with this \citep{Moncuquet2020,Zank2022}. This new property is certainly related to the plasma $\beta$ (ratio of thermodynamic pressure to magnetic pressure) which is often close to unity at 1AU but smaller than one at $\sim 0.1$\,AU. A recent solar wind study at 1\,AU and low $\beta$, reveals the singular role played by fast magneto-acoustic waves \citep{Zhao2022b}: it is shown that this part of turbulence is isotropic with $\alpha = -3/2$. Interestingly, this is a feature that we will demonstrate analytically in the context of fast magneto-acoustic wave turbulence. Since the intensity of the mean magnetic field is expected to be stronger near the Sun, it is natural to think that the wave turbulence regime can provide a relevant description of the young solar wind. 

\subsection{Wave turbulence in MHD}
In incompressible MHD, the Alfvén wave turbulence theory involves three-wave interactions and leads to a strong anisotropy with a cascade only in the direction perpendicular to the uniform magnetic field ${\bf B_0} = B_0 {\bf e_\parallel}$ \citep{Galtier2000,GaltierChandran2006}. The energy spectrum, which is an exact solution of the equations, scales in the simplest case as $k_\perp^{-2}$, and it is expected that MHD turbulence becomes strong at small perpendicular scales \citep{Meyrand2016}. This regime is expected in the solar corona \citep{Bigot2008c,Rappazzo2007}, observed in the Jupiter magnetosphere \citep{Saur2002}, but not in the solar wind. 

A theory of compressible turbulence for MHD is much more difficult to derive because, in particular, one has to deal with three waves: Alfv\'en (A), fast (F) and slow (S) magneto-acoustic waves (see \cite{Galtier2001} for a study of the resonance conditions). A first theory has been proposed in the limit of small $\beta$ and where the main nonlinear mechanism considered is the resonance scattering of (high frequencies) A and F-waves on (low frequencies) S-waves \citep{Kuznetsov2001}. Therefore, this theory involves nonlocal interactions in time scale (or frequency). 
The author found anisotropic spectra for each type of wave, and in particular for F-waves, which is not compatible with isotropic spectra found in observations of the solar wind \citep{Zhao2022b} or in direct numerical simulations \citep{Cho2002,Makwana2020}. 
A second theory has been proposed, in the small $\beta$ limit, but where the S-waves contribution is neglected \citep{Chandran2005}. However, an additional non-physical assumption has also been made on the mass density which has been taken to be constant (it will not be the case in this paper). To be consistent with the energy conservation, the momentum equation was then artificially modified. As a result, the initial equations used are not the original MHD equations, which limits the significance of the predictions. 
Finally, a third theory has been proposed, again in the small $\beta$ limit, to describe three-wave interactions between A, F and S-waves, in the presence of extra terms to model collisionless damping \citep{Chandran2008}. The presence of this effect justifies the absence of three-wave interactions involving only S-waves. However, the complexity of the equations does not allow for exact predictions, especially for F-waves for which an anisotropic spectrum is expected (the reduced case involving only F-waves is mentioned but reference is made to \cite{Chandran2005} where, as explained above, the predictions prove to be limited). 

The aim of this paper is to present a self-consistent and pedagogical theory for compressible MHD turbulence in the small $\beta$ limit, where A and S-waves are neglected (thus retaining only local time-scale interactions, with high frequency fluctuations). Note that this type of approximation has been used for incompressible Hall MHD to derive a theory of wave turbulence where left and right polarized waves have been studied separately \citep{Galtier2006}; the direct numerical simulations show that indeed, at main order, the dynamics between the two types of fluctuations is decoupled \citep{Meyrand2018}. 
It is believed that the exact results obtained with this sub-system can serve as a basis for a better understanding of compressible MHD turbulence, which is a very complex subject. This complexity is underlined by recent direct numerical simulations of subsonic MHD turbulence where a wide range of situations is found, depending in particular on the forcing used. For example, it is shown that F and S-waves become non-negligible compared to A-waves when a compressible forcing is applied instead of an incompressible one \citep{Andres2017,Makwana2020,Gan2022}. The role of these compressible waves can certainly be increased if, instead, a wave forcing is applied as it is usually done in wave turbulence \citep{LeReun2020}. Moreover, the frequency-wavenumber spectra reveal that at small $\beta$ the fluctuations corresponding to S-waves are limited to low frequencies, leaving a large frequency domain for a dynamics driven possibly by fast magneto-acoustic wave turbulence \citep{Andres2017,Brodiano2021}. 

We will see in this paper that the complexity of the kinetic equations of fast magneto-acoustic wave turbulence are relatively limited. This apparent simplicity is linked to the semi-dispersive nature of F-waves with the dispersion relation $\omega \propto k = \sqrt{k_\perp^2 + k_\parallel^2}$. A priori, this leads to some analytical difficulties for the asymptotic closure because we are dealing with three-wave interactions and, in this case, the resonance condition corresponds to collinear wave vectors (note that this constraint does not exist for four-wave interactions -- see e.g. the case of gravitational wave turbulence \citep{Galtier2017}). It turns out that this problem is similar to acoustic wave turbulence \citep{Zakharov1970} for which it was shown that the uniformity of the asymptotic development is broken in one or two dimensions, but can eventually be restored in three dimensions \citep{Benney1966,Newell1971,Lvov1997}. This mathematical property fully justifies the use of the derived wave kinetic equations in this paper, even if a slight correction might be necessary, which can take the form of a broadening of the resonance loci (in our case, rays).

\subsection{The case of semi-dispersive waves}
\label{semidis}

The theory of wave turbulence describes a sea of random waves interacting in a weakly nonlinear manner \citep{GaltierCUP2022}. The great achievement of this theory is the discovery of the existence of a natural asymptotic closure induced by the separation of time scales between the linear and nonlinear times \citep{Benney1966,Benney1967}. It is natural because it does not assume anything about the statistical distributions of the field such as joint Gaussianity \citep{Newell2001}. This represents a breakthrough compared to previous work where the usual procedure was to invoke an {\it ad hoc} statistical assumption in order to close the hierarchy of moment equations \citep{Hasselmann1962}. A necessary ingredient for this success that uses the multiple scale method, is a sufficient degree of decoupling of the initial correlations by the linear response of the system. In simple terms, the reason of the closure is that the cumulant (or moment) evolution separates into two processes. On short time scale, of the order of the wave period (which is $\mathcal{O} (1)$), there is a phase mixing which leads to the decoupling of the correlations initially present and to a statistics that is close to Gaussianity, as expected from the central limit theorem. On a longer time scale (for three-wave interactions, it is $\mathcal{O} (1/\epsilon^2)$, with $\epsilon$ the amplitude of the waves), the nonlinear coupling -- weak over short time -- becomes non-negligible because of the resonance mechanism. This coupling leads to a regeneration of the cumulants via the product of lower order cumulants. It is these contributions that are at the origin of the energy transfer mechanism. 

\begin{figure}
\center{\includegraphics[width=1\linewidth]{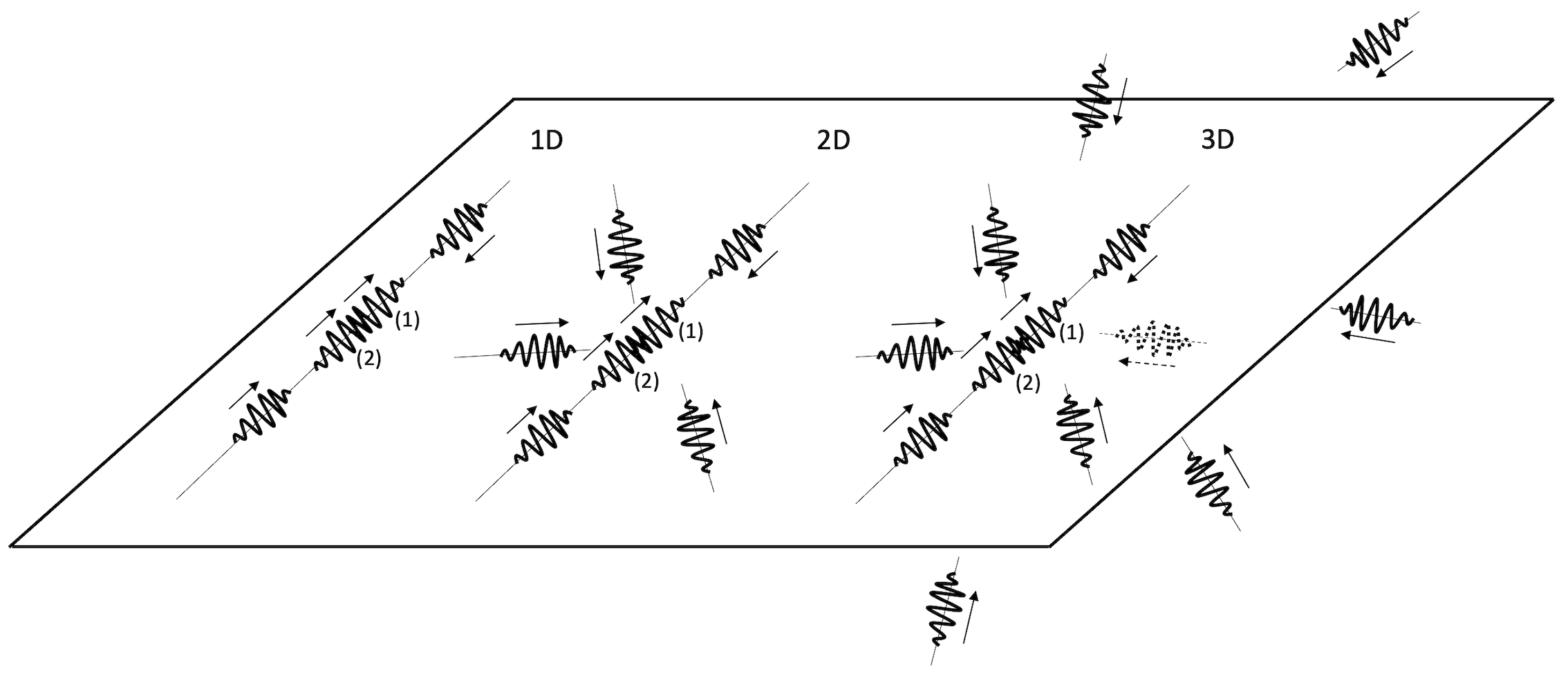}}
\caption{Propagation of wave packets of amplitude $\epsilon$, with $\omega \propto k$, in 1D (left), 2D (middle) and 3D (right). In 1D, the (nonlinear) interaction between wave packets (1) and (2) becomes quickly strong because they are moving in the same direction at the same speed, and thus the initial correlations between the phases of the wave packets persist; collisions between wave packet (1) and those propagating in the opposite direction, which carry statistically independent information, cannot change the situation; therefore, turbulence cannot be weak. In 2D, the number of collisions between wave packet (1) and the others moving in different directions is higher than in 1D but still not enough to change the conclusion. In 3D, this number is sufficient to randomize the phases of wave packet (1) and turbulence can eventually be weak. In pure compressible MHD at $\beta \ll 1$, this explanation applies well for F-waves for which $\omega \propto k$ (semi-dispersive wave). For S-waves where $\omega \propto k_\parallel$ (non-dispersive wave), whatever the dimension, the situation reduces to the 1D case with propagation of wave packets along the strong uniform magnetic field; therefore, turbulence of S-waves cannot be weak \red{and can produce shocks}.}
\label{Fig1}
\end{figure}

It is often thought that the dispersive nature of the waves is a necessary ingredient for achieving an asymptotic closure. The argument is that for non-dispersive waves, all disturbances travel at the same speed and therefore initial correlations between the phases of the waves persist, whereas for dispersive waves any initial correlations are quickly lost as different waves travel with different speeds. However, several comments must be made to nuance this statement. 
First, the analysis is done for three-wave interactions. In this case, the resonance condition leads to rays in Fourier space along which correlations can be preserved. For four-wave interactions, the situation is different because the solutions of the resonance condition are not necessarily confined along rays \citep{Nazarenko2011,Galtier2017,Hassaini2019}. 
Second, for three-wave interactions, the loss of correlation depends on the dimension of the problem. It was shown by \cite{Benney1966} that in one dimension, a natural closure for acoustic wave turbulence is indeed not possible because the initial correlation is preserved, which leads to an energy transfer on a time scale shorter than $\mathcal{O} (1/\epsilon^2)$. Physically, we know that for one-dimensional compressible flow, shocks are formed in a finite time. 
What is true in one  dimension, is not necessarily true in a multidimensional space. Indeed, the fact that many wave packets carrying statistically independent information pass through a given direction can lead to a state close to Gaussianity. Therefore, the central limit theorem can again be operative and a natural closure  occur. However, these phenomenological arguments must be checked carefully. This has been done for acoustic wave turbulence in two-dimension by \cite{Newell1971} and in three-dimension by \cite{Lvov1997}. The conclusion is that in two-dimension the uniformity of the development seems difficult to achieve because of the fast growth of secular terms, however, in three-dimension the growth is much weaker (logarithmic) and a natural closure may be obtained. Although a complete demonstration (involving higher order terms) is still lacking, the study of \cite{Lvov1997} gives the expression of the renormalized frequency needed to restore the uniformity of the development, as well as the generalized kinetic equation which differs from the original equation derived by \cite{Zakharov1970} by the $\delta$ functions which are replaced by Lorenz functions. Interestingly, this generalized equation has the same Kolmogorov-Zakharov solution as the original but, in addition, it allows local angular transfer between adjacent rays. 

The problem studied here, which involves semi-dispersive waves, is similar to acoustic wave turbulence. Therefore, one can assume that the previous development for tridimensional acoustic wave turbulence can be directly applied here. In other words, we will assume that a given wave packet travelling in a fixed direction is crossed by a sufficient number of F-wave packets (carrying statistically independent information) to break its initial correlation. This number is sufficient in three-dimension, but not in two-dimension. This physically explains why the uniformity of the asymptotic development can be preserved for three-dimensional F-wave turbulence, and why the Kolmogorov-Zakharov spectrum that will be derived is indeed a relevant solution of the problem. In Figure \ref{Fig1}, a phenomenological interpretation of this discussion is given with collisions between semi-dispersive wave packets in one (1D), two (2D) and three-dimension (3D).

\subsection{Phenomenology of compressible MHD wave turbulence} \label{sec14}

A recent study carried out at 1 AU \citep{Zhao2022b} reveals that when the plasma $\beta$ is small, solar wind turbulence can be composed of Alfv\'enic fluctuations following the critical balance phenomenology, and of fast magneto-acoustic fluctuations with an isotropic $k^{-3/2}$ energy spectrum. This observation can be understood with the help of simple phenomenological arguments. 
Assuming the existence of a relatively strong uniform magnetic field (written ${\bf b_0}$ in velocity units) and considering the small $\beta$ limit, we obtain the Alfv\'en time 
\be
\tau_A \sim \frac{1}{\omega_A} \sim \frac{1}{k_\parallel b_0} ,
\ee
with $k_\parallel$ the wavenumber component along ${\bf b_0}$, and the fast magneto-acoustic time 
\be
\tau_F \sim \frac{1}{\omega_F} \sim \frac{1}{k b_0} .
\ee
If we assume that the dynamics is mainly governed by Alfvén waves, an anisotropic cascade develops (whatever the regime, weak or strong) with energy mainly located at $k_\perp \gg k_\parallel$. Then, the nonlinear time reads
\be
\tau_{NL} \sim \frac{1}{k b_\ell}  \sim \frac{1}{k_\perp b_\ell} ,
\ee
where $b_\ell$ represents the fluctuations of the magnetic field at a given length scale $\ell$ (for simplicity, we assume equipartition between the velocity and the magnetic field fluctuations). We deduce the following time ratios 
\be
\chi^A = \frac{\tau_A}{\tau_{NL}} \sim \frac{k_\perp b_\ell}{k_\parallel b_0} \quad \rm{and} \quad 
\chi^F =  \frac{\tau_F}{\tau_{NL}} \sim \frac{b_\ell}{b_0}  . 
\ee
If the Alfv\'enic fluctuations follow the critical balance regime as often claimed \citep{Horbury2008}, then $\chi^A \sim 1$ and necessarily we have $\chi^F \ll 1$, which is synonymous with weak F-wave turbulence. Therefore, at any location where the critical balance regime is observed for A-waves (strong wave turbulence), if the plasma is compressible and $\beta$ small, we should find the regime of (weak) F-wave turbulence. Note that signatures of the coexistence of strong and weak wave turbulence in a plasma have already been observed in three-dimensional direct numerical simulations of incompressible Hall MHD where left and right circularly polarized waves are present, the former being in the strong turbulence regime and the latter in the weak turbulence regime \citep{Meyrand2018}. 

Using phenomenological arguments for three-wave interactions, we can also find a prediction for the energy spectrum corresponding to fast magneto-acoustic wave turbulence. We introduce the mean rate of energy transfer (or energy flux) $\varepsilon$ in the inertial range, the transfer (or cascade) time $\tau_{tr}$ and the isotropic energy spectrum $E_k$ (we anticipate that this turbulence is isotropic) such that
\be
\varepsilon \sim \frac{b^2_\ell}{\tau_{tr}} \sim \frac{kE_k}{\omega_F \tau_{NL}^2} \sim \frac{k^2 E_k b^2_\ell}{b_0} \sim \frac{k^3 E^2_k}{b_0} ,
\ee
which gives the one-dimensional energy spectrum
\be
E_k \sim \sqrt{b_0 \varepsilon} k^{-3/2} .
\ee
This is the well-known IK isotropic spectrum \citep{Iroshnikov64,Kraichnan65} often cited in incompressible MHD 
\red{(with the same type of phenomenology -- or dimensional analysis -- a $k^{-3/2}$ energy spectrum can also be found in acoustic wave turbulence \citep{Zakharov1970})}.
However, it is known that this weak isotropic turbulence phenomenology is not well adapted to this situation where anisotropy is expected in the presence of Alfv\'en waves \citep{Galtier2000}. This phenomenology is also not relevant for slow magneto-acoustic waves since we cannot build a theory of weak turbulence. In conclusion, the IK phenomenology of weak turbulence is much better suited to fast magneto-acoustic wave turbulence. Note that the IK spectrum is precisely what was observed by \cite{Zhao2022b} in the solar wind. In this paper, I will show that this isotropic spectrum is in fact an exact solution of fast magneto-acoustic wave turbulence.

\subsection{Plan of the paper}
This paper is organized as follows. In section 2, I present the leading order compressible MHD equations and the compressible Els\"asser fields. In section 3, the wave amplitude equations for the canonical variables are derived. In section 4, I introduce the wave turbulence formalism and derive the kinetic equations for fast magneto-acoustic wave turbulence. In section 5, the properties of the kinetic equations are given with the detailed conservation of energy and momentum. I derive the exact stationary solutions (Kolmogorov-Zakharov spectra), find the locality domain, the sign of the flux and the Kolmogorov constant. In the last section, a conclusion is proposed with a discussion on the relevance of this wave turbulence theory for the solar wind.

\section{Compressible MHD}\label{Sec2}
\subsection{Leading order equations}
Neglecting the dissipative (and forcing) terms, the three-dimensional compressible MHD equations write \citep{GaltierCUP2016}
\be
{\partial \rho \over \partial t} + \nabla \cdot (\rho \uu) = 0 \, , \label{mhd1}
\ee
\be
\rho \left({\partial \uu \over \partial t} + \uu \cdot \nabla \uu \right) = - \nabla P + 
{1 \over \mu_0}(\nabla \times \bb) \times \bb \, , \label{mhd2}
\ee
\be
{\partial \bb \over \partial t} = \nabla \times (\uu \times \bb) \, , \label{mhd3}
\ee
\be
\nabla \cdot \bb = 0 \, , \label{mhd4}
\ee
where $\rho$ is the mass density, $\uu$ the velocity, $P$ the pressure, $\mu_0$ the permeability of free space and $\bb$ the magnetic field. Hereafter, we will consider small mass density fluctuations $\rho_1$ over a uniform density $\rho_0$, namely $\rho \equiv \rho_0 + \rho_1$ with $\rho_1 \ll \rho_0$. We will also neglect the pressure  compared to the magnetic pressure (small $\beta$ limit). We introduce the normalized magnetic field $\bbs \equiv \bb / \sqrt{\mu_0 \rho_0}$ and a uniform (normalized) magnetic field along the parallel direction $\bbs_0 = b_0 {\bf e_\parallel}$ such that $\vert \bbs \vert \equiv b \ll b_0$. 
Under these considerations, the system (\ref{mhd1})--(\ref{mhd3}) reads at leading order
\be
{\partial \rho_1 \over \partial t} + \rho_0 \nabla \cdot \uu = - \nabla \cdot (\rho_1 \uu) \, , \label{mhd1b}
\ee
\be
{\partial \uu \over \partial t} - b_0 \left(\partial_\parallel \bbs - \nabla b_\parallel \right)
= - (\uu \cdot \nabla) \uu - \nabla \left(\frac{\bbs^2}{2} \right) + (\bbs \cdot \nabla) \bbs + \frac{b_0}{\rho_0} \rho_1 \left( \partial_\parallel \bbs - \nabla b_\parallel \right)
\, , \label{mhd2b}
\ee
\be
{\partial \bbs \over \partial t} - b_0 \partial_\parallel \uu + \bbs_0 (\nabla \cdot \uu) 
=  (\bbs \cdot \nabla) \uu - (\uu \cdot \nabla) \bbs - \bbs (\nabla \cdot \uu) \, , \label{mhd3b}
\ee
where the quadratic nonlinear contributions are written in the right hand side and the linear terms in the left hand side. 

The primary vector fields will be decomposed into toroidal ($\psi^{u,b}$), poloidal ($\phi^{u,b}$) and compressible ($\xi$) scalar fields in the following manner
\ba
\uu &=& \nabla \times (\psi^u {\bf e_\parallel}) + \nabla \times (\nabla \times (\phi^u {\bf e_\parallel}) ) + \nabla \xi \, , \\
\bbs &=& \nabla \times (\psi^b {\bf e_\parallel}) + \nabla \times (\nabla \times (\phi^b {\bf e_\parallel}) ) \, .
\ea
Since shear-Alfv\'en waves will be filtered out, we impose $\psi^u = \psi^b = 0$. Then, in Fourier space, we obtain the decomposition
\ba
\hat \uu(\kk) \equiv \hat \uu_k &=& \hat \phi^u_k (k^2 {\bf e_\parallel} - \kpa \kk) + i \hat \xi_k \kk \, , \\
\hat \bbs(\kk) \equiv \hat \bbs_k &=& \hat \phi^b_k (k^2 {\bf e_\parallel} - \kpa \kk) \, ,
\ea
where the symbol $\hat .$ means the Fourier transform and $\kk$ is a wavevector.

\subsection{Compressible Els\"asser variables}
The linearization of (\ref{mhd1})--(\ref{mhd3}) gives 
\ba
{\partial \hat \rho_k \over \partial t} &=& - i \rho_0 \kk \cdot \hat \uu_k \, ,\\
{\partial \hat \uu_k \over \partial t} &=& i \kpa b_0 {\hat \bbs}_k - i b_0 \kk {\hat b}_{\parallel k} \, , \\
{\partial \hat \bbs_k \over \partial t} &=& i \kpa b_0 \hat \uu_k - i \bbs_0 (\kk \cdot \hat \uu_k) \, ,
\ea
which becomes with the decomposition (to simplify the notation, we write $\hat \rho_k \equiv \hat {\rho_1}_k$)
\ba
{\partial \hat \rho_k \over \partial t} &=& \rho_0 k^2 \hat \xi_k \, , \label{Lin1} \\
{\partial \hat \xi_k \over \partial t} &=& - b_0 \kpe^2 \hat \phi^b_k \, , \label{Lin2} \\  
{\partial \hat \phi^u_k \over \partial t} &=& i b_0 \kpa \hat \phi^b_k \, , \label{Lin3} \\
{\partial \hat \phi^b_k \over \partial t} &=& i b_0 \kpa \hat \phi^u_k + b_0 \hat \xi_k \, . \label{Lin4} 
\ea
It is straightforward to show that the dispersion relation is (with our convention $\omega > 0$) 
\be
\omega_k = b_0 k \, .
\ee
From equations (\ref{Lin1})--(\ref{Lin4}), one finds the canonical variables $A_k^s$ (the compressible Els\"asser fields)
\be
A^s (\kk) \equiv A_k^s \equiv \frac{k^2 \kpe}{\kpa} \left( \hat \phi^u_k - s \frac{\kpa}{k} \hat \phi^b_k \right) \, , \label{CV}
\ee
with $s=\pm$ the directional polarity. Interestingly, we find the linear relationships
\be
\hat \xi_k = i \frac{\kpe^2}{\kpa} \hat \phi^u_k \, ,  \label{link1}
\ee
and 
\be 
\hat \rho_k = \frac{\rho_0}{b_0} \kpe^2 \hat \phi^b_k \, . \label{link2}
\ee
This means that the canonical variables have several writings. The choice (\ref{CV}) seems however the most natural since its form is similar to the incompressible Els\"asser fields that involve only $\uu$ and $\bbs$ \citep{GaltierCUP2016}. Another interesting comment is that very often the parallel component of the magnetic field is used to evaluate the compressibility at MHD scales \citep{Zank2022}. Relation (\ref{link1}) demonstrates that here it is not a good proxy.

\subsection{Energy and momentum conservation} \label{sec23}
We assume that the field components have a zero mean value. Then, at leading order and in the small $\beta$ limit, the energy conservation reads
\be
E = \frac{1}{2} \rho_0 \left\langle \uu^2 + \bbs^2 \right\rangle \, ,
\ee
where $\langle  \rangle$ means the spatial average (hereafter, $\langle \rangle$ will be also used as the ensemble average). Note that the fact that the mass density field $\rho_1$ does not appear explicitly in this formula does not mean that it has no effect on the (nonlinear) dynamics. 
In Fourier space, the energy becomes
\ba
E(\kk) &=& \frac{1}{2} \rho_0 \left\langle \hat \uu_k \cdot \hat \uu_k^* + \hat \bbs_k \cdot \hat \bbs_k^* \right\rangle 
= \frac{1}{2} \rho_0 \left\langle \kpe^2 k^2 \left( \vert \hat \phi^u_k \vert^2 + \vert \hat \phi^b_k \vert^2 \right) + k^2 \vert \hat \xi_k \vert^2 \right\rangle  \nonumber \\
&=&  \frac{1}{4} \rho_0 \left\langle \vert A_k^+ \vert^2 + \vert A_k^- \vert^2 \right\rangle \, .
\ea

As it will be proved later, the second invariant is the momentum (or cross-helicity)
\be
H = \frac{1}{2} \rho_0 \left\langle \uu \cdot \bbs \right\rangle \, . 
\ee
In Fourier space, it reads
\ba
H(\kk) &=& \frac{1}{2} \rho_0 \left\langle \hat \uu_k \cdot \hat \bbs_k^* + \hat \bbs_k \cdot \hat \uu_k^* \right\rangle 
= \frac{1}{2} \rho_0 \left\langle \kpe^2 k^2 \left[\hat \phi^u_k (\hat \phi^b_k)^* + (\hat \phi^u_k)* \hat \phi^b_k \right] \right\rangle \nonumber \\
&=& - \frac{1}{4} \rho_0 \frac{\kpa}{k} \left\langle \vert A_k^+ \vert^2 - \vert A_k^- \vert^2 \right\rangle \, .
\ea

\section{Fundamental equation}\label{Sec3}
With the introduction of the canonical variables, one can derive the equation for the F-wave amplitude variation. Its form is
\be
{\partial A_k^s \over \partial t} + i s \omega_k A_k^s = {\cal N}_k \, ,
\ee
with ${\cal N}_k$ the nonlinear contribution in spectral space. A little calculation leads to the following expressions
\be
{\partial \hat \phi^u_k \over \partial t} = i b_0 \kpa \hat \phi^b_k + 
\ee
$$
\frac{i \kpa}{2 \kpe^2 k^2} \int_{\mathbb{R}^6} \left[ \frac{p^2 q^2 \kpe^2}{\ppa \qpa} (\ppeb \cdot \qpeb) \hat \phi_p^u \hat \phi_q^u + 2 \kpa \ppa q^2 (\pp \cdot \qq - \ppa^2) \hat \phi_p^b \hat \phi_q^b \right] \delta_{\kk,\pp\qq} \dd \pp \dd \qq \, , 
$$
and 
\be
{\partial \hat \phi^b_k \over \partial t} = \frac{i b_0 k^2}{\kpa} \hat \phi^u_k + \frac{i}{\kpe^2} \int_{\mathbb{R}^6} \frac{ \ppe^2 q^2}{\qpa} 
\left[ \pp \cdot \qq + \qpe^2-\ppa \qpa \right] \hat \phi_p^b \hat \phi_q^u
\delta_{\kk,\pp\qq} \dd \pp \dd \qq \, , 
\ee
where $\delta_{\kk,\pp\qq} \equiv \delta(\kk-\pp-\qq)$. To derive these expressions, we have used relations (\ref{link1}) and (\ref{link2}). The canonical variables can be introduced by noticing the relations
\ba
\hat \phi^u_k &=& \frac{\kpa}{2\kpe k^2} \sum_s A_k^s \, ,\\
\hat \phi^b_k &=& -\frac{1}{2\kpe k} \sum_s s A_k^s \, .
\ea
We obtain
\be
{\partial \hat \phi^u_k \over \partial t} = i b_0 \kpa \hat \phi^b_k + 
\ee
$$
\frac{i \kpa}{8 \kpe^2 k^2} \int_{\mathbb{R}^6} \sum_{s_p s_q} 
\left[ \frac{\kpe^2}{\ppe \qpe} (\ppeb \cdot \qpeb) + \frac{2 s_p s_q \kpa \ppa q}{p\ppe \qpe} (\pp \cdot \qq - \ppa^2) \right] 
A_p^{s_p} A_q^{s_q} \delta_{\kk,\pp\qq} \dd \pp \dd \qq \, , 
$$
and 
\be
{\partial \hat \phi^b_k \over \partial t} = \frac{i b_0 k^2}{\kpa} \hat \phi^u_k - \frac{i}{4 \kpe^2} \int_{\mathbb{R}^6} \sum_{s_p s_q} \frac{s_p \ppe}{p\qpe}
\left[ \pp \cdot \qq + \qpe^2-\ppa \qpa \right] A_p^{s_p} A_q^{s_q} \delta_{\kk,\pp\qq} \dd \pp \dd \qq \, . 
\ee
Finally, a combination of the two last expressions gives
\be
{\partial A_k^s \over \partial t} + i s \omega_k A_k^s = 
\ee
$$
\frac{i}{8 \kpe} \int_{\mathbb{R}^6} \sum_{s_p s_q} 
\left[ \frac{\kpe^2}{\ppe \qpe} (\ppeb \cdot \qpeb) + \frac{2s_ps_q\kpa \ppa q}{p\ppe \qpe} (\pp \cdot \qq - \ppa^2)
+ \frac{2s s_p k \ppe}{p\qpe} (\pp \cdot \qq + \qpe^2-\ppa \qpa) \right] 
$$
$$A_p^{s_p} A_q^{s_q} \delta_{\kk,\pp\qq} \dd \pp \dd \qq \, ,$$
which can also be written in a symmetric form
\be
{\partial A_k^s \over \partial t} + i s \omega_k A_k^s = 
\ee
$$
\frac{i}{8 \kpe} \int_{\mathbb{R}^6} \sum_{s_p s_q} 
\left[ \frac{\kpe^2}{\ppe \qpe} (\ppeb \cdot \qpeb) 
+ \frac{\kpa s_ps_q}{\ppe \qpe pq} \left( \ppa q^2 (\pp \cdot \qq - \ppa^2) + \qpa p^2 (\pp \cdot \qq - \qpa^2) \right) \right.
$$
$$\left. + \frac{sk}{\ppe \qpe pq} \left( s_p \ppe^2q (\ppeb \cdot \qpeb + \qpe^2) + s_q \qpe^2 p (\ppeb \cdot \qpeb + \ppe^2) \right) \right] 
A_p^{s_p} A_q^{s_q} \delta_{\kk,\pp\qq} \dd \pp \dd \qq \, .$$
The form of the wave amplitude equation is non-trivial, however, several simplifications are possible. First, we shall use the interaction representation and consider a wave of small amplitude ($0 < \epsilon \ll 1$)
\be
A_k^s = \epsilon a_k^s e^{-is\omega_kt} \, ,
\ee
which leads to
\be
{\partial a_k^s \over \partial t} =
\ee
$$\frac{i \epsilon}{8 \kpe} \int_{\mathbb{R}^6} \sum_{s_p s_q} \left[ \frac{\kpe^2}{\ppe \qpe} (\ppeb \cdot \qpeb) 
+ \frac{\kpa s_ps_q}{\ppe \qpe pq} \left( \ppa q^2 (\pp \cdot \qq - \ppa^2) + \qpa p^2 (\pp \cdot \qq - \qpa^2) \right) \right.
$$
$$
\left. +\frac{sk}{\ppe \qpe pq} \left( s_p \ppe^2q (\ppeb \cdot \qpeb + \qpe^2) + s_q \qpe^2 p (\ppeb \cdot \qpeb + \ppe^2) \right) \right] 
a_p^{s_p} a_q^{s_q} e^{i\Omega_{k,pq} t} \delta_{\kk,\pp\qq} \dd \pp \dd \qq \, ,
$$
with $\Omega_{k,pq} \equiv s\omega_k - s_p\omega_p - s_q\omega_q$. Second, we will anticipate the consequence of the resonance condition that the kinetic equations must satisfy. This condition 
\ba
sk &=& s_pp + s_qq \, , \\
\kk &=& \pp + \qq \, ,
\ea
leads to collinear wavevectors and thus to the relations $\pp \cdot \qq = s_p s_q pq$ and $\ppeb \cdot \qpeb = s_p s_q \ppe \qpe$. We also have $\ppe^2\qpa^2=\qpe^2 \ppa^2$. With this information, the wave amplitude equation reduces to
\be
{\partial a_k^s \over \partial t} = 
\frac{i \epsilon}{8 \kpe} \int_{\mathbb{R}^6} \sum_{s_p s_q} \left[ s_p s_q \kpe^2 + \frac{\kpa}{pq} \left( \ppa q^2 + \qpa p^2 \right) \right.
\ee
$$\left. + \frac{sks_ps_q}{pq} \left( sk\ppe\qpe + s_p \ppe^2q  + s_q \qpe^2 p \right) \right] 
a_p^{s_p} a_q^{s_q} e^{i\Omega_{k,pq} t} \delta_{\kk,\pp\qq} \dd \pp \dd \qq \, .
$$
A last simplification can be made by introducing $\theta$, the angle between $\kk$ and ${\bf e_\parallel}$, and thus, the relations $\kpa = k \cos \theta_k$, $p_\parallel=s s_p p \cos \theta_k$ and $q_\parallel=s s_q q \cos \theta_k$. We also have $\kpe = k \sin \theta_k$, $p_\perp = p \sin \theta_k$ and $q_\perp = q \sin \theta_k$. Since the wavevectors $\kk$, $\pp$ and $\qq$ are collinear, we arrive at
\be
{\partial a_k^s \over \partial t} = i \epsilon \int_{\mathbb{R}^6} \sum_{s_p s_q} \left(\frac{1 + 2 \sin^2 \theta_k}{8 \sin \theta_k}\right) k s_p s_q
a_p^{s_p} a_q^{s_q} e^{i\Omega_{k,pq} t} \delta_{\kk,\pp\qq} \dd \pp \dd \qq \, . \label{Aequa}
\ee
The presence of $\sin \theta_k$ in the denominator cannot lead to a divergence because by definition the canonical variable is null for $\theta_k = 0$. 
Note that in the small $\beta$ limit, the displacement vectors of the fast waves are almost transverse to ${\bf e_\parallel}$, which is the opposite limit $\theta_k \simeq \pi/2$ \citep{GaltierCUP2016}.
Equation (\ref{Aequa}) is the fundamental equation of our problem and its form is classical for three-wave interactions. As expected, we see that the nonlinear terms are of order $\epsilon$. This means that weak nonlinearities will only change the amplitude of the F-waves slowly over time. 
The nonlinearities contain an exponentially oscillating term that is essential for the asymptotic closure. Indeed, the theory of wave turbulence deals with variations of spectral densities at very large times, i.e. for a nonlinear transfer time much larger than the F-wave period: in other words, we have a time scale separation between the fast oscillations of the waves (due to the phase variations in the exponential) and the slow variations of the wave amplitudes. As a consequence, most of the nonlinear terms are destroyed and only a few of them, the resonance terms, for which $\Omega_{k,pq} =0$, survive \citep{Benney1966,Benney1967,Benney1967b}. 
From equation (\ref{Aequa}), we finally see that, contrary to incompressible MHD, there are no exact solutions to the nonlinear problem. The origin of such a difference is that in incompressible MHD the nonlinear term implies Alfv\'en waves moving only in opposite directions \citep{Galtier2000} whereas in purely compressible MHD this constraint does not exist (we have a summation over $s_p$ and $s_q$). In other words, if one type of wave is not present in incompressible MHD then the nonlinear term cancels out whereas in the present problem this is not the case. 

Note that it is legitimate to wonder whether a derivation based on the other variables ($\hat \rho_k, \hat \xi_k$) could lead to another expression that would invalidate our initial choice of canonical variables. A similar calculation based on the equations (\ref{Lin1})-(\ref{Lin2}) including the nonlinear terms leads to the same expression as (\ref{Aequa}), thus proving the consistency of the present derivation. 
Hereafter, we shall introduce the following variable $c_k^s \equiv a_k^s /\sqrt{\omega_k}$ (linked to the action) that will facilitate the derivation of the kinetic equations. We obtain
\be
{\partial c_k^s \over \partial t} = i \epsilon \int_{\mathbb{R}^6} \sum_{s_p s_q} L_{-kpq}^{-ss_ps_q} 
c_p^{s_p} c_q^{s_q} e^{i\Omega_{k,pq} t} \delta_{\kk,\pp\qq} \dd \pp \dd \qq \, , \label{Aequa2}
\ee
where the interaction coefficient
\be
L_{kpq}^{ss_ps_q} \equiv  \sqrt{b_0}  \left(\frac{1 + 2 \sin^2 \theta_k}{8 \sin \theta_k}\right) s_p s_q \sqrt{kpq} \, ,
\ee
satisfies the following symmetries
\ba
L_{kpq}^{ss_{p}s_{q}} &=& L_{kqp}^{ss_{q}s_{p}} \, , \\
L_{0pq}^{ss_{p}s_{q}} &=& 0 \, , \\
L_{-k-p-q}^{ss_{p}s_{q}} &=& L_{kpq}^{ss_{p}s_{q}} \, , \\
L_{kpq}^{-s-s_{p}-s_{q}} &=& L_{kpq}^{ss_{p}s_{q}} \, , \\
ss_{q} L_{qpk}^{s_{q}s_{p}s} &=& L_{kpq}^{ss_{p}s_{q}} \, , \\
ss_{p} L_{pkq}^{s_{p}ss_{q}} &=& L_{kpq}^{ss_{p}s_{q}} \, .
\ea

\section{Derivation of the kinetic equations}\label{Sec3}
We now move on to a statistical description. We use the ensemble average $\langle \rangle$ and define the following spectral correlators (cumulants) for homogeneous turbulence (we will also assume $\langle \aks \rangle=0$)
\ba
\langle \aks \akss \rangle &=& q^{ss'}_{kk'}(\kk,\kk') \delta(\kk+\kk') \, , \label{stat1} \\
\langle \aks \akss \aksss \rangle &=& q^{ss's''}_{kk'k''}(\kk,\kk',\kk'') \delta(\kk+\kk'+\kk'') \, , \label{stat2}  \\
\langle \aks \akss \aksss \akssss \rangle &=& q^{ss's''s'''}_{kk'k''k'''}(\kk,\kk',\kk'',\kk''') \delta(\kk+\kk'+\kk''+\kk''') \label{stat3} \\
&+& q^{ss'}_{kk'}(\kk,\kk') q^{s''s'''}_{k''k'''}(\kk'',\kk''') \delta(\kk+\kk') \delta(\kk''+\kk''') \nonumber\\
&+& q^{ss''}_{kk''}(\kk,\kk'') q^{s's'''}_{k'k'''}(\kk',\kk''') \delta(\kk+\kk'') \delta(\kk'+\kk''') \nonumber\\
&+& q^{ss'''}_{kk'''}(\kk,\kk''') q^{s's''}_{k'k''}(\kk',\kk'') \delta(\kk+\kk''') \delta(\kk'+\kk'') \nonumber \, .
\ea
From the fundamental equation (\ref{Aequa2}), we get
\ba
\frac{\partial \langle \aks \akss \rangle}{\partial t} &=& 
\left\langle \frac{\partial \aks}{\partial t} \akss \right\rangle + \left\langle \aks \frac{\partial \akss}{\partial t} \right\rangle \label{2ndo} \\
&=& i \epsilon \int_{\RR^{6}} \sum_{s_{p} s_{q}}  L_{-kpq}^{-ss_{p}s_{q}} \langle \akss \aksp \aksq \rangle 
e^{i \Okpq t} \del \dd \pp \dd \qq \nonumber \\
&+& i \epsilon \int_{\RR^{6}} \sum_{s_{p} s_{q}}  L_{-k'pq}^{-s's_{p}s_{q}} \langle \aks \aksp \aksq \rangle 
e^{i\Okpqp t} \delp \dd \pp \dd \qq \nonumber \, . 
\ea
At the next order we have
\ba
\frac{\partial \langle \aks \akss \aksss \rangle}{\partial t} &=& 
\left\langle \frac{\partial \aks}{\partial t} \akss \aksss \right\rangle 
+ \left\langle \aks \frac{\partial \akss}{\partial t} \aksss \right\rangle 
+ \left\langle \aks \akss \frac{\partial \aksss}{\partial t} \right\rangle \label{cumul3} \\
&=& i \epsilon \int_{\RR^{6}} \sum_{s_{p} s_{q}}  L_{-kpq}^{-ss_{p}s_{q}} \langle \akss \aksss \aksp \aksq \rangle 
e^{i \Okpq t} \del \dd \pp \dd \qq \nonumber \\
&+& i \epsilon \int_{\RR^{6}} \sum_{s_{p} s_{q}}  L_{-k'pq}^{-s's_{p}s_{q}} \langle \aks \aksss \aksp \aksq \rangle 
e^{i \Okpqp t} \delp \dd \pp \dd \qq \nonumber \\
&+& i \epsilon \int_{\RR^{6}} \sum_{s_{p} s_{q}}  L_{-k''pq}^{-s''s_{p}s_{q}} \langle \aks \akss \aksp \aksq \rangle 
e^{i \Okpqpp t} \delpp \dd \pp \dd \qq \nonumber \, .
\ea
Here we face the classic problem of closure: a hierarchy of statistical equations of increasingly higher order emerges (see discussion in Section \ref{semidis}). In contrast to the strong turbulence regime, in the weak wave turbulence regime we can use the scale separation in time to achieve a natural closure of the system \citep{Benney1966}. Expressions (\ref{stat1})--(\ref{stat3}) are introduced into equation (\ref{cumul3})
\ba
&&\frac{\partial q^{ss's''}_{kk'k''}(\kk,\kk',\kk'')}{\partial t} \delta(\kk+\kk'+\kk'') = \label{q4t1} \\
&& i \epsilon \int_{\RR^{6}} \sum_{s_{p} s_{q}}  L_{-kpq}^{-ss_{p}s_{q}} 
\left[ q^{s's''s_{s_{p}}s_{s_{q}}}_{k'k''pq}(\kk',\kk'',\pp,\qq) \delta(\kk'+\kk''+\pp+\qq) \right. \nonumber \\
&&+ q^{s's''}_{k'k''}(\kk',\kk'') q^{s_{p}s_{q}}_{pq}(\pp,\qq) \delta(\kk'+\kk'') \delta(\pp+\qq) \nonumber \\
&&+ q^{s's_{p}}_{k'p}(\kk',\pp) q^{s''s_{q}}_{k''q}(\kk'',\qq) \delta(\kk'+\pp) \delta(\kk''+\qq) \nonumber \\
&&\left. + q^{s's_{q}}_{k'q}(\kk',\qq) q^{s''s_{p}}_{k''p}(\kk'',\pp) \delta(\kk'+\qq) \delta(\kk''+\pp) \right] e^{i \Okpq t} \del \dd \pp \dd \qq \nonumber \\
&&+ \, i \epsilon \int_{\RR^{6}} \Big\{ (\kk,s) \leftrightarrow (\kk',s') \Big\} \dd \pp \dd \qq \nonumber \\
&&+ \, i \epsilon \int_{\RR^{6}} \Big\{ (\kk,s) \leftrightarrow (\kk'',s'') \Big\} \dd \pp \dd \qq \, ,  \nonumber
\ea
where the last two lines correspond to the exchange at the notation level between $\kk$, $s$ in the expanded expression and $\kk'$, $s'$ (penultimate line), then $\kk''$, $s''$ (last line). 

We are now going to integrate expression (\ref{q4t1}) both on $\pp$ and $\qq$, and on time, by considering a long integrated time compared to the reference time (the F-wave period). The presence of several Dirac functions leads to the conclusion that the second term on the right (in the main expression) gives no contribution since it corresponds to $k=0$ for which the interaction coefficient is null. It is a property of statistical homogeneity. The last two terms on the right (always in the main expression) lead to a strong constraint on wavevectors $\pp$ and $\qq$ which must be equal to $-\kk'$ or $-\kk''$. For the fourth-order cumulant, the constraint is much less strong since only the sum of $\pp$ and $\qq$ is imposed. A consequence is that for long times this term will not contribute to the non-linear dynamics \citep{GaltierCUP2022}. Finally, for long times the second-order cumulants are only relevant when the associated polarities have different signs. In order to understand this, it is necessary to go back to the definition of the moment, $\langle \Aks \Akss \rangle = \epsilon^{2} \langle a_k^s a_{k'}^{s'} \rangle \exp(-i(s\omega_{k}+s'\omega_{k'})t)$, from which we see that in the limit of large time a non-zero contribution is possible for homogeneous turbulence ($\kk=-\kk'$) only if $s=-s'$ (then the coefficient of the exponential is cancelled). We finally get
\ba
&&q^{ss's''}_{kk'k''}(\kk,\kk',\kk'') \delta(\kk+\kk'+\kk'') = i \epsilon \Delta(\Omega_{kk'k''}) \delta(\kk+\kk'+\kk'') \label{q4t2} \\
&& \left\{ \left[ L_{-k-k'-k''}^{-s-s'-s''} + L_{-k-k''-k'}^{-s-s''-s'} \right]  q^{s''-s''}_{k''-k''}(\kk'',-\kk'') q^{s'-s'}_{k'-k'}(\kk',-\kk') \right. \nonumber \\
&& +\left[ L_{-k'-k-k''}^{-s'-s-s''} + L_{-k'-k''-k}^{-s'-s''-s} \right]  q^{s''-s''}_{k''-k''}(\kk'',-\kk'') q^{s-s}_{k-k}(\kk,-\kk) \nonumber \\
&& \left. + \left[ L_{-k''-k'-k}^{-s''-s'-s} + L_{-k''-k-k'}^{-s''-s-s'} \right]  q^{s-s}_{k-k}(\kk,-\kk) q^{s'-s'}_{k'-k'}(\kk',-\kk') \right\}\, ,  \nonumber
\ea
with
\be
\Delta(\Omega_{kk'k''}) = \int_0^{t\gg1/\omega} e^{i \Omega_{kk'k''}t^\prime} dt^\prime 
=  {e^{i \Omega_{kk'k''}t} - 1 \over i \Omega_{kk'k''} } \, .
\ee
We can now write without ambiguity: $q^{s-s}_{k-k}(\kk,-\kk) = q^s_k(\kk)$.  Using the symmetry relations of the interaction coefficient, we obtain
\ba
&&q^{ss's''}_{kk'k''}(\kk,\kk',\kk'') \delta(\kk+\kk'+\kk'') = - 2 i \epsilon \Delta(\Omega_{kk'k''}) \delta(\kk+\kk'+\kk'') \label{q4t3} \\
&& \left[ L_{kk'k''}^{ss's''} q^{s''}_{k''}(\kk'') q^{s'}_{k'}(\kk') + L_{k'kk''}^{s'ss''} q^{s''}_{k''}(\kk'') q^{s}_{k}(\kk) 
+ L_{k''k'k}^{s''s's} q^{s}_{k}(\kk) q^{s'}_{k'}(\kk') \right] \, ,  \nonumber
\ea
and then
\ba
&&q^{ss's''}_{kk'k''}(\kk,\kk',\kk'') \delta(\kk+\kk'+\kk'') = - 2 i \epsilon \Delta(\Omega_{kk'k''}) \delta(\kk+\kk'+\kk'') \label{q4t4} \quad \\
&& L_{kk'k''}^{ss's''}  \left[ q^{s''}_{k''}(\kk'') q^{s'}_{k'}(\kk') + ss' q^{s''}_{k''}(\kk'') q^{s}_{k}(\kk) + ss'' q^{s}_{k}(\kk) q^{s'}_{k'}(\kk') \right] \, . 
 \nonumber
\ea
The effective long time limit (which introduces irreversibility) gives us (Riemann-Lebesgue's lemma)
\be
\Delta(x) \to \pi \delta(x) + i {\cal P} (1/x) \, , 
\ee
with ${\cal P}$ the principal value integral. 

The so-called kinetic equation is obtained by injecting expression (\ref{q4t4}) in the long time limit, into equation (\ref{2ndo}) and integrating on $\kk'$ (with the relation $q^{-s}_{-k}(-\kk)=q^{s}_{k}(\kk)$)
\ba
&&\frac{\partial q^{s}_{k}(\kk)}{\partial t} = 2 \epsilon^{2} \int_{\RR^{6}} \sum_{s_{p} s_{q}}  \vert L_{-kpq}^{-ss_{p}s_{q}} \vert^2
(\pi \delta(\Omega_{-kpq}) + i {\cal P} (1/\Omega_{-kpq}))  e^{i \Okpq t} \del  \nonumber \\
&& s_{p}s_{q} \left[ s_{p}s_{q} q^{s_{q}}_{q}(\qq) q^{s_{p}}_{p}(\pp) - ss_{q} q^{s_{q}}_{q}(\qq) q^{s}_{k}(\kk) 
- ss_{p} q^{s}_{k}(\kk) q^{s_{p}}_{p}(\pp) \right]
\dd \pp \dd \qq \nonumber \\
&+& 2 \epsilon^{2} \int_{\RR^{6}} \sum_{s_{p} s_{q}}  \vert L_{kpq}^{ss_{p}s_{q}} \vert^2
(\pi \delta(\Omega_{kpq}) + i {\cal P} (1/\Omega_{kpq})) e^{i\Omega_{kpq} t} \delta_{kpq}  \\
&& s_{p}s_{q} \left[ s_{p}s_{q} q^{s_{q}}_{q}(\qq) q^{s_{p}}_{p}(\pp) + ss_{q} q^{s_{q}}_{q}(\qq) q^{s}_{k}(\kk) 
+ ss_{p} q^{s}_{k}(\kk) q^{s_{p}}_{p}(\pp) \right]
\dd \pp \dd \qq \nonumber \, . 
\ea
By changing the sign of the (dummy) variables $\pp$ and $\qq$ of integration, and the associated polarities, the principal values are eliminated. Using the symmetries of the interaction coefficient, we finally arrive at the following expression after simplification
\ba
\frac{\partial q^{s}_{k}(\kk)}{\partial t} &=& 
\frac{\pi \epsilon^{2} b_0}{16} \int_{\RR^{6}} \sum_{s_{p} s_{q}} \left(\frac{1 + 2 \sin^2 \theta_k}{\sin \theta_k}\right)^2 kpq
\delta(\Omega_{kpq})
\delta(\kk+\pp+\qq) \label{KE1} \\
&& s_{p}s_{q}  \left[ s_{p}s_{q} q^{s_{q}}_{q}(\qq) q^{s_{p}}_{p}(\pp) + ss_{q} q^{s_{q}}_{q}(\qq) q^{s}_{k}(\kk) 
+ ss_{p} q^{s}_{k}(\kk) q^{s_{p}}_{p}(\pp) \right] \dd \pp \dd \qq  \, . \nonumber 
\ea
Expression (\ref{KE1}) is the kinetic equation of fast magneto-acoustic wave turbulence. The presence of the small parameter $\epsilon \ll 1$ means that the amplitude of the quadratic non-linearities is weak and that, consequently, the characteristic time over which we place ourselves to measure these effects is of the order of $1/\epsilon^2$ (the reference time being the wave period $1/\omega$). 

\section{Properties of F-wave turbulence}\label{Sec4}
\subsection{Detailed conservation}
A remarkable property verified by the kinetic equation (\ref{KE1}) is the detailed conservation of the invariants (energy and momentum for three-wave interactions). To demonstrate this result, the kinetic equation must be rewritten for the polarized energy spectrum
\be
e^s(\kk) \equiv \omega_{k} q^{s}_{k}(\kk) \, .
\ee
One notices in particular that, $e^s(\kk)=e^{-s}(-\kk)$. After a few manipulations, we get
\ba
\frac{\partial e^s(\kk)}{\partial t} &=& 
\frac{\pi \epsilon^{2}}{16 b_0^2} \int_{\RR^{6}} \sum_{s_{p} s_{q}}  \left(\frac{1 + 2 \sin^2 \theta_k}{\sin \theta_k}\right)^2
\delta(\Omega_{kpq}) \delta(\kk+\pp+\qq) \quad \quad \label{KE2} \\
&& s \omega_{k} 
\left[ \frac{s\omega_{k}}{e^s(\kk)} + \frac{s_{p} \omega_{p}}{e^{s_{p}}(\pp)} + \frac{s_{q} \omega_{q}}{e^{s_{q}}(\qq)} \right] 
e^{s}(\kk) e^{s_{p}}(\pp) e^{s_{q}}(\qq) \dd \pp \dd \qq \, . \nonumber 
\ea
By considering the integral in $\kk$ of the total energy spectrum, $E(\kk) \equiv \sum_s e^s(\kk)$, we find
\ba
&&\frac{\partial \int_{\RR^{3}} E(\kk) d\kk}{\partial t} = 
\frac{\pi \epsilon^{2}}{16 b_0^2} \int_{\RR^{9}} \sum_{s s_{p} s_{q}}  \left(\frac{1 + 2 \sin^2 \theta_k}{\sin \theta_k}\right)^2
\delta(\Omega_{kpq}) \delta(\kk+\pp+\qq) \quad \quad \label{KE3}  \\
&& (s \omega_k +s_p \omega_p +s_q \omega_q) 
\left[ \frac{s\omega_{k}}{e^s(\kk)} + \frac{s_{p} \omega_{p}}{e^{s_{p}}(\pp)} + \frac{s_{q} \omega_{q}}{e^{s_{q}}(\qq)} \right] 
e^{s}(\kk) e^{s_{p}}(\pp) e^{s_{q}}(\qq) \dd \kk \dd \pp \dd \qq = 0 \, .  \nonumber
\ea
This means that energy is conserved by triadic interaction: the redistribution of energy takes place within a triad satisfying the resonance condition. 

The second invariant is the momentum also called cross-helicity in MHD (see section \ref{sec23}). The polarized cross-helicity spectrum is defined as
\be
h^s(\kk) \equiv \frac{\kpa}{k} e^s(\kk) \, .
\ee
After a few manipulations, we find
\ba
\frac{\partial h^s(\kk)}{\partial t} &=& 
\frac{\pi \epsilon^2}{16} \int_{\RR^{6}} \sum_{s_{p} s_{q}}  \left(\frac{1 + 2 \sin^2 \theta_k}{\sin \theta_k}\right)^2
\delta(\Omega_{kpq}) \delta(\kk+\pp+\qq) \quad \quad \label{KE2} \\
&& \frac{kpq}{\kpa\ppa\qpa} s \kpa
\left[ \frac{s\kpa}{h^s(\kk)} + \frac{s_{p} \ppa}{h^{s_{p}}(\pp)} + \frac{s_{q} \qpa}{h^{s_{q}}(\qq)} \right] 
h^{s}(\kk) h^{s_{p}}(\pp) h^{s_{q}}(\qq) \dd \pp \dd \qq \, . \nonumber 
\ea
By introducing the total cross-helicity spectrum $H(\kk) \equiv \sum_s s h^s(\kk)$, we obtain
\ba
&&\frac{\partial \int_{\RR^{3}} H(\kk) d\kk}{\partial t} = 
\frac{\pi \epsilon^{2}}{16} \int_{\RR^{9}} \sum_{s s_{p} s_{q}}  \left(\frac{1 + 2 \sin^2 \theta_k}{\sin \theta_k}\right)^2
\delta(\Omega_{kpq}) \delta(\kk+\pp+\qq) \quad \quad \label{KE2} \\
&& \frac{kpq}{\kpa\ppa\qpa} (\kpa+\ppa+\qpa)
\left[ \frac{s\kpa}{h^s(\kk)} + \frac{s_{p} \ppa}{h^{s_{p}}(\pp)} + \frac{s_{q} \qpa}{h^{s_{q}}(\qq)} \right] 
h^{s}(\kk) h^{s_{p}}(\pp) h^{s_{q}}(\qq) \dd \kk \dd \pp \dd \qq = 0 \, . \nonumber 
\ea
This means that cross-helicity is conserved by triadic interaction and its redistribution takes place within a triad satisfying the resonance condition. 

\subsection{Angular anisotropic spectra}
For discussion purposes, it is best to rewrite the kinetic equation for energy as follows
\ba
\frac{\partial e^s(\kk)}{\partial t} &=& \left(\frac{1 + 2 \sin^2 \theta_k}{\sin \theta_k}\right)^2
\frac{\pi \epsilon^{2}}{16} \int_{\RR^{6}} \sum_{s_{p} s_{q}}  
\delta(\Omega_{kpq}) \delta(\kk+\pp+\qq) \quad \quad \label{Kx1} \\
&& s k 
\left[ sk e^{s_p}(\pp) e^{s_q}(\qq) + s_p p e^s(\kk) e^{s_q}(\qq) + s_q q e^s(\kk) e^{s_p}(\pp) \right] 
 \dd \pp \dd \qq \, , \nonumber 
\ea
where the dependence in $\theta_k$ has been placed outside the integral since it depends only on $\kk$. Before deriving the exact power law solutions of the kinetic equation, we will use a property deduced from the resonant condition. We know (and have already used) that the wavevectors are aligned, which means that the energy cascade develops along rays and thus each spectra within the integral has the same angular dependence. However, the coefficient in front of the integral depends on $\theta_k$: in particular, the smaller $\theta_k$ is, the larger the coefficient is, and therefore the smaller the transfer time is. This property is compatible with the phenomenology introduced in section \ref{sec14}: the cascade tends to be stronger along the parallel direction. 
Note that for acoustic wave turbulence, there is no such  angular dependence, so the problem is more isotropic than in MHD \citep{Zakharov1970}. 

From the previous remark, we introduce the reduced spectrum 
\be
k^2 e^s(\kk)= k^2 e^s(k,\theta_k,\phi_k) = f(\theta_k,\phi_k) E^s(k) = f(\theta_k,\phi_k) E^s_k \, .
\ee 
The function $f(\theta_k,\phi_k) \ge 0$ depends on the initial condition: once given, it will not change by the turbulence cascade because there is no redistribution of energy in $\theta_k$ or $\phi_k$. This leads to
\ba
\frac{\partial E^s_k}{\partial t} &=& \left(\frac{1 + 2 \sin^2 \theta_k}{\sin \theta_k}\right)^2 f(\theta_k,\phi_k)
\frac{\pi \epsilon^{2}}{16} \int_{\RR^{6}} \sum_{s_{p} s_{q}}  
\delta(\Omega_{kpq}) \delta(\kk+\pp+\qq) \quad \quad \label{Kx2} \\
&& \frac{s k}{p^2q^2} 
\left[ sk^3 E^{s_p}_p E^{s_q}_q + s_p p^3 E^s_k E^{s_q}_q + s_q q^3 E^s_k E^{s_p}_p \right] 
 \dd \pp \dd \qq \, . \nonumber 
\ea
Since there is no angular dependence in the integral except for the delta function, we can perform angular averaging and use the following relationship \citep{ZLF92}
\be
\langle \delta(\kk+\pp+\qq) \rangle_{\rm angle} = 
\int_{\RR^{4}} \delta(\kk+\pp+\qq) \dd \cos \theta_p \dd \cos \theta_q \dd \phi_p \dd \phi_q = \frac{1}{2kpq} \, .
\ee
We obtain
\be \label{510}
\frac{\partial E^s_k}{\partial t} = 
\ee
$$
\frac{\pi \epsilon^{2} K_{\theta,\phi}}{32b_0} \int_{\Delta_\perp} \sum_{s_{p} s_{q}}  
\delta(sk+s_pp+s_qq)\frac{s}{pq} 
\left[ sk^3 E^{s_p}_p E^{s_q}_q + s_p p^3 E^s_k E^{s_q}_q + s_q q^3 E^s_k E^{s_p}_p \right]  \dd p \dd q \, ,  
$$
with $\Delta_\perp$ the integration domain (infinitely long band) and by definition 
\be
K_{\theta,\phi} = \left(\frac{1 + 2 \sin^2 \theta_k}{\sin \theta_k}\right)^2 f(\theta_k,\phi_k) \, .
\ee
We recall that by construction, the canonical variables (\ref{CV}) cancel for $\theta_k=0$ (as does the spectrum), so we will not consider this limit in the following. The exact solutions can now be derived using the Zakharov transform. It will provide power law spectra at given angles $(\theta_k, \phi_k)$. Therefore, this problem is anisotropic but of a very special type because it does not imply different power laws in parallel and perpendicular directions as is usually found in plasma physics \citep{Galtier2003b,Galtier2006,Galtier2014,Galtier2015}. An exception is the case of incompressible MHD \citep{Galtier2000} where no cascade is possible along the parallel direction (a function $f(\kpa)$ is then introduced whose form depends on the initial condition). Note that a similar qualitative dependence in $\theta_k$ (but not in $\phi_k$) has been reported by \cite{Chandran2005} but, as explained in the introduction, the compressible MHD equations have been artificially modified to satisfy energy conservation with a constant mass density. Unlike \cite{Chandran2005}, the mass density spectrum can be predicted here (see below). 

From expression (\ref{510}) we can deduce the kinetic equations for energy and momentum, which are respectively
\be \label{kinE}
\frac{\partial E_k}{\partial t} = 
\frac{\pi \epsilon^{2} K_{\theta,\phi}}{128 b_0} \int_{\Delta_\perp} \sum_{ss_{p} s_{q}}  
\delta(sk+s_pp+s_qq)\frac{s}{pq} 
\left[ sk^3 \left( E_p E_q + H_p H_q / \cos^2 \theta_k \right) \right.
\ee
$$
\left. + s_p p^3 \left(E_k E_q + H_k H_q/ \cos^2 \theta_k \right)
+ s_q q^3 \left(E_k E_p + H_k H_p/ \cos^2 \theta_k \right)
\right]  \dd p \dd q \, ,  
$$
and 
\be \label{kinH}
\frac{\partial H_k}{\partial t} = 
\frac{\pi \epsilon^{2} K_{\theta,\phi}}{128 b_0} \int_{\Delta_\perp} \sum_{ss_{p} s_{q}}  
\delta(sk+s_pp+s_qq)\frac{1}{pq} 
\left[ k^3 \left( E_p H_q + E_q H_p \right) \right.
\ee
$$
\left. + p^3 \left(E_k H_q + E_q H_k \right) + q^3 \left(E_k H_p + E_p H_k \right)
\right]  \dd p \dd q \, ,  
$$
with (see section \ref{sec23}) $E_k \equiv E_k^+ + E_k^-$ and $H_k \equiv - (k_\parallel / k) (E_k^+ - E_k^-)$.

\subsection{Kolmogorov-Zakharov spectra}
In this section, we shall derive the exact power-law solutions of the kinetic equations (\ref{kinE})--(\ref{kinH}). We introduce 
\be \label{ECK}
E_k \equiv C_E k^x \quad {\rm and} \quad H_k \equiv C_H k^y \, .
\ee
and the dimensionless wavenumbers $\tilde p \equiv p/k$ and $\tilde q \equiv q/k$. 
$C_E$ and $C_H$ are two constants such that $C_E \in \RR^+$ and $C_H \in \RR$. It leads to
\be \label{kinE2}
\frac{\partial E_k}{\partial t} = 
\frac{\pi \epsilon^{2} K_{\theta,\phi}}{128 b_0} k^{2+2x} \int_{\Delta_\perp} \sum_{ss_{p} s_{q}}  
\delta(s+s_p \tilde p+s_q \tilde q)\frac{s}{\tilde p \tilde q} 
\left[ C_E^2 \left( s {\tilde p}^x {\tilde q}^x + s_p {\tilde p}^3 {\tilde q}^x + s_q {\tilde p}^x {\tilde q}^3 \right) \right.
\ee
$$
+ \left. C_H^2 \left( s{\tilde p}^y {\tilde q}^y + s_p{\tilde p}^3 {\tilde q}^y + s_q{\tilde p}^y {\tilde q}^3 \right) k^{2y-2x} / \cos^2 \theta_k 
 \right] \dd {\tilde p} \dd {\tilde q} \, ,  
$$
and 
\be \label{kinH2}
\frac{\partial H_k}{\partial t} = 
\frac{\pi \epsilon^{2} K_{\theta,\phi}}{128 b_0} k^{x+y+2} C_E C_H \int_{\Delta_\perp} \sum_{ss_{p} s_{q}}  
\delta(s+s_p \tilde p+s_q \tilde q)\frac{1}{\tilde p \tilde q} 
\left[ {\tilde p}^x {\tilde q}^y + {\tilde p}^y {\tilde q}^x \right.
\ee
$$
\left. + {\tilde p}^3 {\tilde q}^y + {\tilde p}^3 {\tilde q}^x + {\tilde p}^y {\tilde q}^3 + {\tilde p}^x {\tilde q}^3 
\right]  \dd \tilde p \dd \tilde q \, .
$$
The Zakharov transform \citep{ZLF92} consists in splitting the kinetic equations into three parts and applying to two of them the following change of variables
\be
\tilde p \to \frac{1}{\tilde p} \, , \quad \tilde q \to \frac{\tilde q}{\tilde p} \, , \label{TZ1}
\ee
and
\be
\tilde p \to \frac{\tilde p}{\tilde q} \, , \quad \tilde q \to \frac{1}{\tilde q} \, . \label{TZ2}
\ee
We obtain for the energy
\be \label{kinE3}
\frac{\partial E_k}{\partial t} = 
\frac{\pi \epsilon^{2} K_{\theta,\phi}}{384 b_0} k^{2+2x} \int_{\Delta_\perp} \sum_{ss_{p} s_{q}}  
\delta(s+s_p \tilde p+s_q \tilde q)\frac{1}{\tilde p \tilde q} 
\ee
$$
\left\{ C_E^2 \left[ s \left( s {\tilde p}^x {\tilde q}^x + s_p {\tilde p}^3 {\tilde q}^x + s_q {\tilde p}^x {\tilde q}^3 \right) 
+ s_p {\tilde p} \left( s_p {\tilde p}^{-2x} {\tilde q}^x + s {\tilde p}^{-3-x} {\tilde q}^x + s_q {\tilde p}^{-3-x} {\tilde q}^3 \right) \right. \right.
$$
$$
\left.  + s_q {\tilde q} \left( s_q {\tilde p}^{x} {\tilde q}^{-2x} + s_p {\tilde p}^3 {\tilde q}^{-3-x} + s {\tilde p}^x {\tilde q}^{-3-x} \right) \right]
$$
$$
+ \frac{k^{2y-2x} C_H^2}{\cos^2 \theta_k}  \left[ s \left( s{\tilde p}^y {\tilde q}^y + s_p {\tilde p}^3 {\tilde q}^y + s_q {\tilde p}^y {\tilde q}^3 \right)  
+ s_p {\tilde p} \left( s_p {\tilde p}^{-2y} {\tilde q}^y + s {\tilde p}^{-3-y} {\tilde q}^y + s_q {\tilde p}^{-3-y} {\tilde q}^3 \right) \right.
$$
$$
+ \left. \left. s_q {\tilde q} \left( s_q {\tilde p}^y {\tilde q}^{-2y} + s_p {\tilde p}^3 {\tilde q}^{-3-y} + s {\tilde p}^y {\tilde q}^{-3-y} \right) 
 \right] \right\} \dd {\tilde p} \dd {\tilde q} \, ,  
$$
which can be written in a compact form as
\be \label{kinE4}
\frac{\partial E_k}{\partial t} = 
\frac{\pi \epsilon^{2} K_{\theta,\phi}}{384 b_0} k^{2+2x} \int_{\Delta_\perp} \sum_{ss_{p} s_{q}}  
\delta(s+s_p \tilde p+s_q \tilde q)
\ee
$$
\left[ C_E^2 {\tilde p}^{x-1} {\tilde q}^{x-1} \left(s+s_p {\tilde p}^{-2x-2} +s_q {\tilde q}^{-2x-2} \right) 
\left(s+s_p {\tilde p}^{3-x} +s_q {\tilde q}^{3-x} \right) \right. 
$$
$$
\left. + \frac{k^{2y-2x}C_H^2}{ \cos^2 \theta_k} {\tilde p}^{y-1} {\tilde q}^{y-1}
\left( s+ s_p{\tilde p}^{-2y-2} + s_q{\tilde q}^{-2y-2} \right) 
\left(s+ s_p{\tilde p}^{3-y} + s_q{\tilde q}^{3-y} \right) \right] 
\dd {\tilde p} \dd {\tilde q} \, .
$$
We can find exact stationary solutions. First, we have $x=y=2$ as a spectrum with zero energy flux: this is the thermodynamic solution for which we have no cascade. The other (more interesting) solution corresponds to $x=y=-3/2$: it is the Kolmogorov-Zakharov spectrum for which the energy flux is finite. 
In the two cases, the helicity spectrum is associated with an energy flux; in other words, its dynamics is driven by the energy cascade. 

The Zakharov transform applied to the momentum (cross-helicity) gives
\be \label{kinH2}
\frac{\partial H_k}{\partial t} = 
\frac{\pi \epsilon^{2} K_{\theta,\phi}}{384 b_0} k^{x+y+2} C_E C_H \int_{\Delta_\perp} \sum_{ss_{p} s_{q}}  
\delta(s+s_p \tilde p+s_q \tilde q)\frac{1}{\tilde p \tilde q} 
\ee
$$
\left[ {\tilde p}^x {\tilde q}^y + {\tilde p}^y {\tilde q}^x + {\tilde p}^3 {\tilde q}^y +{\tilde p}^3 {\tilde q}^x + {\tilde p}^y {\tilde q}^3 + {\tilde p}^x {\tilde q}^3 \right.
$$
$$
+ \tilde p ({\tilde p}^{-x-y} {\tilde q}^y + {\tilde p}^{-x-y} {\tilde q}^x + {\tilde p}^{-3-y} {\tilde q}^y + {\tilde p}^{-3-x} {\tilde q}^x + {\tilde p}^{-3-y} {\tilde q}^3 
+ {\tilde p}^{-3-x} {\tilde q}^3)
$$
$$
\left. + \tilde q ({\tilde p}^x {\tilde q}^{-x-y} + {\tilde p}^y {\tilde q}^{-x-y} + {\tilde p}^{3} {\tilde q}^{-3-y} + {\tilde p}^3 {\tilde q}^{-3-x} + {\tilde p}^y {\tilde q}^{-3-y} 
+ {\tilde p}^x {\tilde q}^{-3-x}) \right]
\dd {\tilde p} \dd {\tilde q} \, ,
$$
which can be written in a compact form as
\be \label{kinH2}
\frac{\partial H_k}{\partial t} = 
\frac{\pi \epsilon^{2} K_{\theta,\phi}}{384 b_0} k^{x+y+2} C_E C_H \int_{\Delta_\perp} \sum_{ss_{p} s_{q}}  
\delta(s+s_p \tilde p+s_q \tilde q)\frac{1}{\tilde p \tilde q} \left( 1+ {\tilde p}^{-x-y-2} + {\tilde q}^{-x-y-2} \right) 
\ee
$$
\left[ {\tilde p}^x {\tilde q}^y \left(1+ {\tilde p}^{3-x} + {\tilde q}^{3-y} \right) 
+  {\tilde p}^y {\tilde q}^x \left(1+ {\tilde p}^{3-y} + {\tilde q}^{3-x} \right) \right]
\dd {\tilde p} \dd {\tilde q} \, . 
$$
We see that in this case no stationary solution is possible. The finite cross-helicity flux solution $x+y=-3$ (first line) gives the coefficient $1+ {\tilde p} + {\tilde q}$ that does cancel on the resonance manifold. For the thermodynamic solution, $x=y=2$, we arrive to the same conclusion. 

In conclusion, the most relevant exact solution for fast magneto-acoustic wave turbulence is the one-dimensional Kolmogorov-Zakharov energy spectrum 
\be \label{KZS}
E_k = C_E k^{-3/2} \, ,
\ee
characterized by a finite energy flux. 
Interestingly, this is the well-know Iroshnikov-Kraichnan (IK) spectrum \citep{Iroshnikov64,Kraichnan65} proposed many years ago for incompressible MHD. This is also the exact solution for acoustic wave turbulence \citep{Zakharov1970}, however, a difference exists between the two problems because, unlike acoustic waves, here the spectrum depends on the angle $\theta_k$ between the wavevector and the direction of the applied magnetic field. As we will see, this anisotropy appears in the amplitude $C_E$ of the energy spectrum, with a modulation of its amplitude. 

As explained above, the constant energy flux solutions leads also to a cross-helicity spectrum $H_k \sim k^{-3/2}$. Using the definition of the canonical variables (\ref{CV}) and relation (\ref{link2}), we find dimensionally 
\be
E^{\rho}_k \sim k^{-3/2} ,
\ee
where $E^{\rho}_k$ is the one-dimensional spectrum of density fluctuations. Although a Kolmogorov-type spectrum in $k^{-5/3}$ is often found at 1\,AU in the solar wind \citep{Chen2014}, near the Sun this spectrum is slightly less steep \citep{Moncuquet2020}. This difference can be interpreted as an evolution of the turbulence regime, from weak to strong as the solar wind expands.

\subsection{Locality condition}
\begin{figure}
\center{\includegraphics[width=1\linewidth]{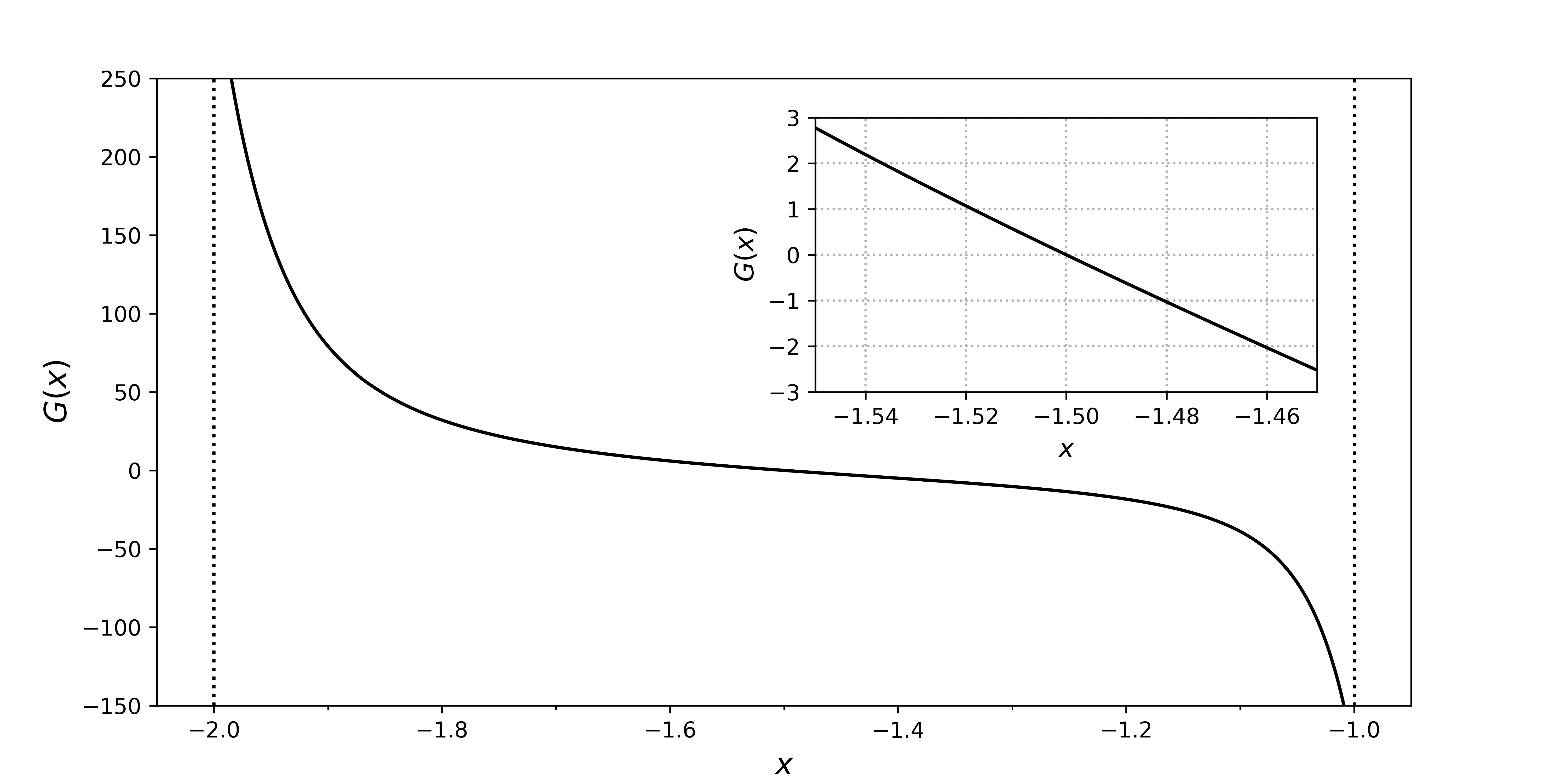}}
\caption{Variation of $G(x)$ for $x \in [-2,-1]$. A divergence of the integral is clearly observed close to $-2$ and $-1$. Inset: as expected, we see that the power law index $x=-3/2$ (Kolmogorov-Zakharov spectrum) cancels the integral.}
\label{fig1}
\end{figure}

The Kolmogorov-Zakharov energy spectrum (\ref{KZS}) found previously is an exact solution of the problem which is only relevant if it satisfies the locality condition. This condition consists in checking the convergence of the integrals in the case of strongly non-local interactions (it will be done in the case $C_H=0$). Physically, the convergence ensures that the solution is independent of the physics at large and small scales where forcing and dissipation are dominant. This calculation must be done {\it before} the application of the Zakharov transformation. Therefore, we consider the expression (hereafter, the small parameter $\epsilon$ is removed since it is a measure of the time scale) 
\be
\frac{\partial E_k}{\partial t} = 
\frac{\pi K_{\theta,\phi} C_E^2 k^{2+2x}}{128 b_0}  \sum_{ss_{p} s_{q}}  \int_{\Delta_\perp} 
\delta(s+s_p \tilde p+s_q \tilde q)\frac{s}{\tilde p \tilde q} 
\left( s {\tilde p}^x {\tilde q}^x + s_p {\tilde p}^3 {\tilde q}^x + s_q {\tilde p}^x {\tilde q}^3 \right)
\dd {\tilde p} \dd {\tilde q} \, ,  
\ee
that we integrate once; one finds
\be
\frac{\partial E_k}{\partial t} = 
\frac{\pi K_{\theta,\phi} C_E^2 k^{2+2x}}{32b_0} \sum_{ss_{p} s_{q}} I^{ss_ps_q}(x) \, ,
\ee
with
\be
I^{ss_ps_q}(x) = \frac{1}{4}\int_0^{+\infty} 
\left[ ({\tilde p}^{x-1} + s s_p{\tilde p}^2)(-ss_q -s_ps_q \tilde p)^{x-1} +ss_q {\tilde p}^{x-1} (s+s_p\tilde p)^2 \right] \dd {\tilde p} \, . 
\ee
For a question of convergence, it is relevant to rewrite this sum of integrals as a single integral. 
\red{First, we apply the following change of variables: $y=1/{\tilde p}$ for $I^{+-+}$, $y=1/({\tilde p}+1)$ for $I^{++-}$, and $y=p$ for $I^{+--}$; we obtain ($s=s_p=s_q$ being not allowed)}
\red{\ba
G(x) &=& \sum_{ss_{p} s_{q}} I^{ss_ps_q}(x) \\
&=& {1 \over 2} \int_0^{1} \left[ (1-y)^{x-1} (y^{x-1}-y^2+2y^{-2x}-2y^{-x-3}) + (1-y)^2 (2y^{-3-x}-y^{x-1}) \right] \dd y \, . \nonumber 
\ea}
\red{Then, we split the integral in two parts ($\int_0^{1/2} + \int_{1/2}^{1}$) and get after modification of the second integral (with a change of variable $y \to 1-y$)}
\ba
&& G(x) = \int_0^{1/2} 
\left[ (1-y)^{x-1} (y^{x-1}-y^2+y^{-2x}-y^{-x-3}) \right. \\
&& \left. + (1-y)^2(y^{-3-x}-y^{x-1}) + y^{x-1} \left[(1-y)^{-2x}-(1-y)^{-x-3}\right] + y^2(1-y)^{-3-x} \right] \dd y \, . \nonumber 
\ea
The condition of convergence must be studied only when $y \to 0$. A detailed calculation leads to the following condition 
\be
-2 < x < -1 \, , 
\ee
which justifies the relevance of the Kolmogorov-Zakharov energy spectrum. As very often, the power law index found for the constant flux solution is placed exactly in the middle of the convergence domain. The variation of $G(x)$ is shown in Figure \ref{fig1}.

\subsection{Direction of the cascade}

\begin{figure}
\center{\includegraphics[width=.9\linewidth]{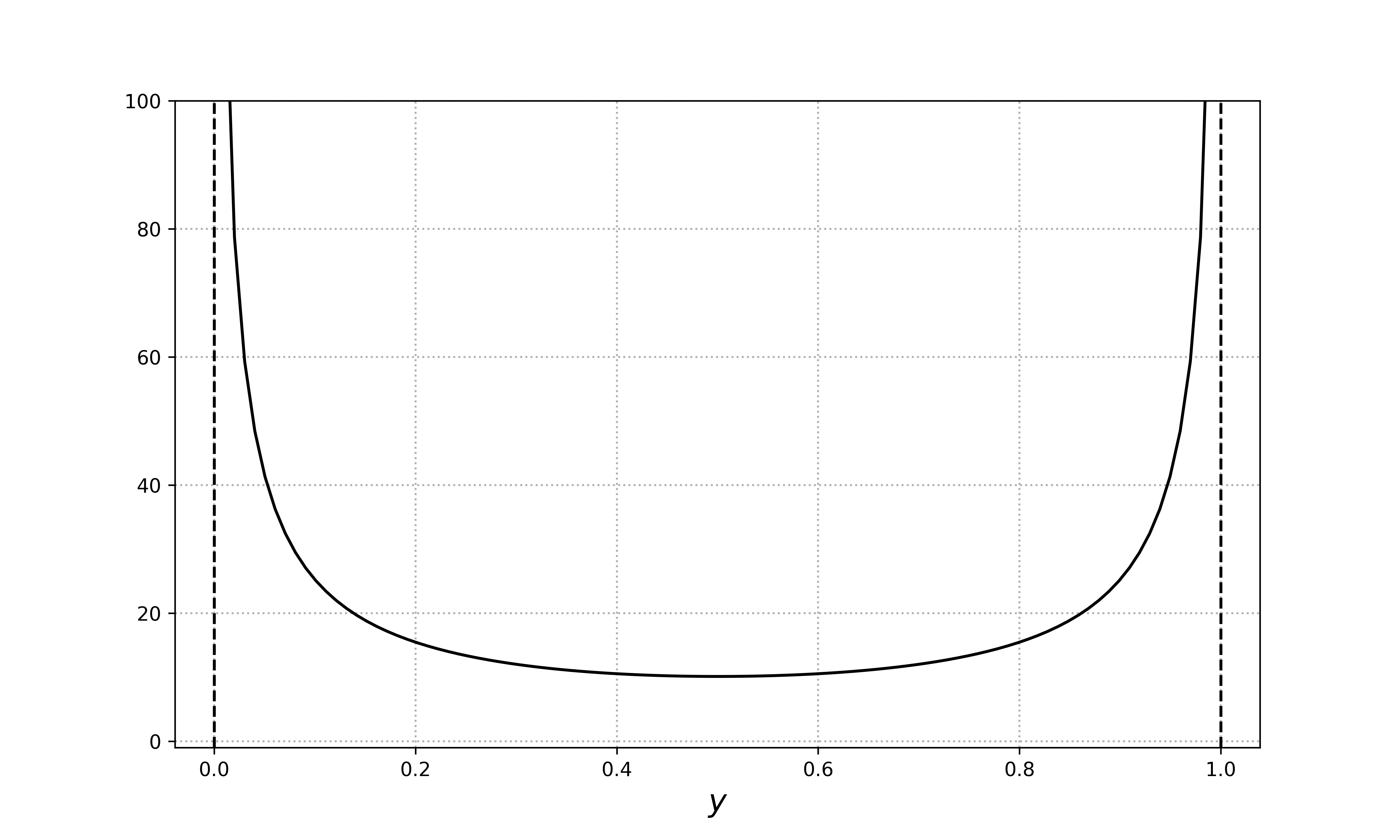}}
\caption{Variation of the integrand of $J$ in expression (\ref{integrand}).}
\label{fig2}
\end{figure}

The next analytical result of this paper is about the direction of the energy cascade. We can prove that this cascade is direct. 
We introduce the isotropic energy flux $\Pi_{k}$ such that 
\be
\frac{\partial E_k}{\partial t} = - \frac{\partial \Pi_k} {\partial k} = \frac{\pi K_{\theta,\phi} C_E^2 k^{2+2x}}{32b_0} I(x)\, , 
\ee
where
\be
I(x) = \frac{1}{12} \int_{\Delta_\perp} \sum_{ss_{p} s_{q}} {\tilde p}^{x-1} {\tilde q}^{x-1} 
\left(s+s_p {\tilde p}^{3-x} +s_q {\tilde q}^{3-x} \right) \left(s+s_p {\tilde p}^{-2x-2} +s_q {\tilde q}^{-2x-2} \right)
\ee
$$
\delta(s+s_p \tilde p+s_q \tilde q)  \dd {\tilde p} \dd {\tilde q} \, .
$$
Here, we use expression (\ref{kinE4}) obtained {\it after} applying the Zakharov transformation. We get
\be
\Pi_k = - \frac{\pi K_{\theta,\phi} C_E^2 k^{3+2x}}{32b_0} \frac{I(x)}{3+2x} \, . 
\ee
The direction of the cascade will be given by the sign of the energy flux when $x=-3/2$ (Kolmogorov-Zakharov spectrum), but in this limit, the numerator and denominator cancel out. The use of L'Hospital's rule leads to the relation
\ba
\lim_{x \to -3/2} \Pi_k \equiv \varepsilon &=& - \frac{\pi K_{\theta,\phi} C_E^2}{32b_0} \lim_{x \to -3/2} \frac{I(x)}{3+2x} \\
&=& - \frac{\pi K_{\theta,\phi} C_E^2}{32b_0} \frac{\p I(x)/\p x\vert_{x=-3/2}}{2}
= \frac{\pi K_{\theta,\phi} C_E^2}{32b_0} J \, ,
\ea
with 
\red{
\be
J = \frac{1}{12} \sum_{ss_{p} s_{q}} J^{ss_ps_q} 
\ee
and
\be
J^{ss_ps_q} = \int_{\Delta_\perp} {\tilde p}^{-5/2} {\tilde q}^{-5/2} 
\left(s+s_p {\tilde p}^{9/2} +s_q {\tilde q}^{9/2} \right) \left(s_p {\tilde p} \ln {\tilde p}  +s_q {\tilde q} \ln {\tilde q} \right)
\ee
$$
\delta(s+s_p \tilde p+s_q \tilde q) \dd {\tilde p} \dd {\tilde q} \, .
$$}
After a few manipulation, $J$ can be written as a one dimensional integral. 
\red{To find this integral, we first use the following change of variables: ${\tilde q} = {\tilde p} +1$ for $J^{++-}$, ${\tilde q} = {\tilde p} -1$ for $J^{+-+}$ and ${\tilde q} = 1-{\tilde p}$ for $J^{+--}$, which leads to
\be
J^{++-} = \int_0^{+\infty} x^{-5/2} (1+x)^{-5/2} \left(1+x^{9/2} - (1+x)^{9/2} \right) \left(x \ln x - (1+x) \ln (1+x) \right) \dd x \, ,
\ee
\be
J^{+-+} = J^{++-}
\ee
and
\be
J^{+--} = - \int_0^{1} x^{-5/2} (1-x)^{-5/2} \left(1-x^{9/2} - (1-x)^{9/2} \right) \left(x \ln x + (1-x) \ln (1-x) \right) \dd x \, .
\ee
Then, we introduce the change of variable $y = 1/(x+1)$ for $J^{++-}$ which becomes equal to $J^{+--}$. This leads  after summation to}
\be \label{integrand}
J = \frac{1}{2} \int_0^1 y^{-5/2} (1-y)^{-5/2} \left(y^{9/2} + (1-y)^{9/2} -1 \right) \left(y \ln y + (1-y) \ln (1-y) \right) \dd y \, .
\ee
The integrand of $J$ is always positive (see Figure \ref{fig2}), therefore the energy flux is positive (because $K_{\theta,\phi} \ge 0$) and the energy cascade direct. 

\subsection{Kolmogorov constant}

The last result of this paper is the numerical evaluation of the universal (Kolmogorov) constant $C_K$ of this problem. From the previous expression and the definition (\ref{ECK}), we deduce the analytical expression 
\be
E_k = \sqrt{\frac{b_0 \varepsilon}{K_{\theta,\phi}}} C_K k^{-3/2} \quad {\rm with} \quad C_K \equiv \sqrt{\frac{32}{\pi J}} \, .
\ee
A numerical evaluation of $J$ (\ref{integrand}) leads to the Kolmogorov constant
\be
C_K \simeq 0.623 \, .
\ee
The one-dimensional energy spectrum is not universal since it depends on $K_{\theta,\phi}$, thus on the anisotropic nature of the system (and also the initial condition). Note that if initially the spectrum is isotropic, then $f(\theta_k,\phi_k) = 4 \pi$ and the expression simplifies. In this case, we can define the Kolmogorov constant as $C'_K = C_K / \sqrt{4 \pi} \simeq 0.176$. 

With the definition of $K_{\theta,\phi}$, the energy spectrum tends to zero when $\theta_k \to 0$, which is consistent with the idea that the contribution of fast waves becomes negligible in this limit. 
However, when a measurement is made in the solar wind, the spectrum may be affected by the change in direction of ${\bf e_\parallel}$ such that only a mean value around a cone of angle $\theta_0 \ll \pi $ is accessible. This effect can be evaluated by introducing $\theta_0$ and the following functions
\ba
g(\theta_k) &=& \frac{\sin \theta_k}{1+2 \sin^2 \theta_k} , \\
\bar g(\theta_0)  &=& \frac{1}{\theta_0} \int_0^{\theta_0} g(\theta_k)  d\theta = \frac{1}{\sqrt{24}\theta_0}
\left( \ln \left( \frac{\sqrt{3}+\sqrt{2}}{\sqrt{3}-\sqrt{2}} \right) - \ln \left( \frac{\sqrt{3}+\sqrt{2} \cos \theta_0}{\sqrt{3}-\sqrt{2} \cos \theta_0} \right) \right) .
\ea
With $\theta_0 = \pi/18$ ($10^o$), we get $\bar g(\pi/18) \simeq 0.085$ and $g(\pi/18)=0.164$ (note that $g(\pi/2)=1/3$). Therefore, we see that $g$ saturates at a relatively high value with $\bar g(\pi/18) / g(\pi/18) \simeq 52\%$. 
This remark can explain the observations where the variation in amplitude of the spectrum does not change very much with the angle $\theta_k$ \citep{Zhao2022b}.

\section{Conclusion and discussion}

In this paper, an analytical theory of wave turbulence is derived for compressible MHD in the small $\beta$ limit for which slow magneto-acoustic waves and Alfv\'en waves are neglected. Then, the nonlinear dynamics is reduced to three-wave interactions between fast magneto-acoustic waves. We find the canonical variables -- the compressible Els\"asser fields: this is a non-trivial combination of the poloidal components of the velocity and magnetic field. These variables are linearly related to the compressible velocity and mass density, respectively. In particular, this means that the parallel component of the magnetic field is not a good proxy to estimate the compressibility. We show that the kinetic equations of wave turbulence possess two quadratic invariants, energy and momentum, which are conserved in detail. However, a relevant exact power-law solution (Kolmogorov-Zakharov spectrum) exists only for the energy: it is the well-known one-dimensional isotropic Iroshnikov-Kraichnan spectrum in $k^{-3/2}$ which finds here a rigorous justification. Interestingly, the mass density spectrum follows also the same scaling. We prove rigorously that this solution is local and corresponds to a direct cascade. The analytical expression of the Kolmogorov constant is also obtained and a numerical estimate is given. 
Unlike acoustic waves \red{\citep{Zakharov1970}}, fast magneto-acoustic wave turbulence is not isotropic in the sense that the amplitude of the spectrum depends on the angle between the wavevector and the direction of the applied uniform magnetic field. 

It is often believed that a theory of wave turbulence is only possible for dispersive waves. The argument is that for non-dispersive waves, all disturbances move at the same speed and therefore initial correlations between the wave phases persist and lead, over a long period of time, to strong turbulence, whereas for dispersive waves, any initial correlations are quickly lost as different waves travel with different speeds. 
However, this statement should be taken with caution for several reasons. 
First, it is implicitly assumed that we have three-wave interactions: in this case, the resonance condition implies solutions along rays, which means that interacting waves are indeed propagating in the same direction. But for four-wave interactions, the situation is different because the solutions of the resonance condition are not necessarily confined along rays (see e.g. elastic waves \citep{Hassaini2019} or gravitational waves \citep{Galtier2017}). 
Second, even for three-wave interactions, one can find exceptions. A well-known example is given by Alfv\'en waves for which $\omega \propto k_\parallel$ and for which nonlinear interactions occur only between waves propagating in opposite directions. Therefore, there is no cumulative effect and a theory of wave turbulence can be developed \citep{Galtier2000}. 
Third, a $\omega \propto k$ relation is non-dispersive only in one dimension. In two or three dimensions, it is semi-dispersive. By semi-dispersive we physically mean that a wave packet moving in a fixed direction interacts with many other wave packets carrying statistically independent information. It turns out that in three dimensions, this number of interactions can be sufficient to break the initial correlation; we then find ourselves in the same situation as for dispersive waves (but with some particularities). 
From a mathematical point of view, it was recognized very early that a theory of acoustic wave turbulence is not feasible in one dimension because the uniformity of the development is not guaranteed \citep{Benney1966}. The same problem seems inevitable in two dimensions but not in three dimensions \citep{Newell1971,Lvov1997}, justifying a posteriori the results obtained by \cite{Zakharov1970}. 
A recent three-dimensional direct numerical simulation shows for the first time the existence of the regime of acoustic wave turbulence with, as expected, an energy spectrum in $k^{-3/2}$ \citep{Kochurin2022}. 
The similarity between acoustic and fast magneto-acoustic waves fully justifies the development of a wave turbulence theory and the use of the exact solutions (Kolmogorov-Zakharov spectrum) as signature of this regime. 
Note that for acoustic waves, a small dispersion is sometimes introduced to justify the existence of the wave turbulence regime. In MHD, such a dispersion can be played naturally by the Hall effect which leads to a correction at small MHD scales \citep{Galtier2006}. \red{In the particular case of three-wave interactions between dispersive kinetic Alfvén waves, an angular distribution of energy is found with a reduction of the cascade along the uniform magnetic field \citep{GaltierCUP2022}.}

Interestingly, recent observations made by PSP reveal a more universal solar wind near the Sun ($\sim 0.1$ AU) than near the Earth in that the power law index found for the kinetic and magnetic energies is $-3/2$ \citep{Chen2020,Shi2021,Zhao2022}. In light of the present study, it is not surprising that at the same time $\beta$ is generally much less than one. For the future, it would be interesting to extend the study by \cite{Zhao2022b} to other data and also check whether the mass density spectrum is consistent with the $-3/2$ power law index \citep{Moncuquet2020,Zank2022} (whereas it seems rather compatible with $-5/3$ at 1AU \citep{Montgomery1987,Coles1989,Hnat2005}). 
It seems also interesting to consider fast magneto-acoustic wave turbulence as a relevant regime to study the heating of the solar corona 
\citep{Galtier1998,Chandran2005}.

\bibliographystyle{jpp}
\bibliography{jpp-biblio}
\end{document}